\documentclass[12pt]{article}
\pdfoutput=1

\usepackage{amssymb,amsmath,amsfonts,eurosym,geometry,ulem,graphicx,caption,color,setspace,sectsty,comment,footmisc,caption,pdflscape,subfigure,array}
\usepackage[numbers]{natbib}
\newtheorem{proposition}{Proposition}

\newtheorem{definition}{Definition}[section]
\newtheorem{property}{Property}[section]
\newtheorem{remark}[proposition]{Remark}
\newtheorem{lemma}{Lemma}[section]

\usepackage{amsmath,mathtools}
\usepackage{amssymb}

\usepackage{dsfont}

\newcolumntype{L}[1]{>{\raggedright\let\newline\\arraybackslash\hspace{0pt}}m{#1}}
\newcolumntype{C}[1]{>{\centering\let\newline\\arraybackslash\hspace{0pt}}m{#1}}
\newcolumntype{R}[1]{>{\raggedleft\let\newline\\arraybackslash\hspace{0pt}}m{#1}}

\makeatletter
\def\blfootnote{\xdef\@thefnmark{}\@footnotetext}
\makeatother

\usepackage{color, colortbl}
\definecolor{LightCyan}{rgb}{0.88,1,1}
\definecolor{Gray}{gray}{0.9}

\begin{document}

\begin{titlepage}
\title{Accountability and Insurance in IoT Supply Chain}
\author{Yunfei Ge \and Quanyan Zhu}
\date{\today}
\maketitle

\begin{abstract}
\noindent\\
Supply chain security has become a growing concern in security risk analysis of the Internet of Things (IoT) systems. Their highly connected structures have significantly enlarged the attack surface, making it difficult to track the source of the risk posed by malicious or compromised suppliers. This chapter presents a system-scientific framework to study the accountability in IoT supply chains and provides a holistic risk analysis technologically and socio-economically. We develop stylized models and quantitative approaches to evaluate the accountability of the suppliers. Two case studies are used to illustrate accountability measures for scenarios with single and multiple agents. Finally, we present the contract design and cyber insurance as economic solutions to mitigate supply chain risks. They are incentive-compatible mechanisms that encourage truth-telling of the supplier and facilitate reliable accountability investigation for the buyer.
\end{abstract}

\blfootnote{Department of Electrical and Computer Engineering, Tandon School of Engineering}
\blfootnote{New York University, Brooklyn, NY, 11201, USA}
\blfootnote{E-mail: \{yg2047, qz494\}@nyu.edu}

\setcounter{page}{0}
\thispagestyle{empty}
\end{titlepage}
\pagebreak \newpage

\section{Introduction} \label{sec:introduction}

Supply chains play a critical role in the security and resilience of IoT systems and affect many users, including small- and medium-sized businesses and government agencies. An attacker can exploit vulnerabilities of a vendor in the supply chain to compromise the IoT system at the end-user. The recent SolarWinds attack is an example of an attack that has resulted in a series of data breaches at government agencies. One seller of the Microsoft Cloud services was compromised by the attacker, allowing the attacker to access the customer data of its resellers. Once the attacker established a foothold in SolarWind's software publishing infrastructure after getting access to SolarWind's Microsoft Office 365 account, he stealthily planted malware into software updates that were sent to the users, which include customers at US intelligence services, executive branch, and military.  

The infamous Target data breach in 2013 is another example of supply-chain attacks. The attacker first broke into Target's main data network through ill-protected HVAC systems. The attacker exploited the vulnerabilities in the monitoring software of the HVAC systems, which shared the same network with the data services. It led to a claimed total loss of \$290 million to data breach-related fees \cite{farris_2015,manworren2016you}. The supply-chain attacks would become increasingly pervasive in IoT systems. Consider a next-generation industrial manufacturing plant equipped with IoT devices that are supported by third-party vendors. The software and the hardware of these devices can be trojanized. As a result, the attacker disrupts the manufacturing plant, which can create a shortage of essential products (e.g., pharmaceutical products, COVID19 vaccines, and gasoline) and lead to grave repercussions in the nation's supply chain. 

\begin{figure}[ht]
\centering
 \includegraphics[width=.8\linewidth]{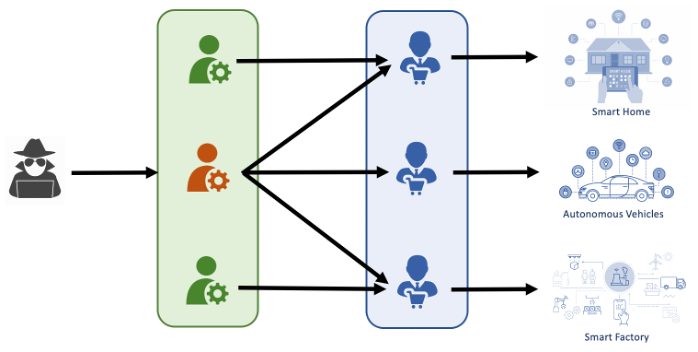}
 \caption{Supply-chain attacks: An attacker first attacks a vendor, who sells the users compromised products. They act as Trojans inside the user's system and stealthily manipulate it. }
\label{fig:supplychainattack}
\end{figure}

Risk-based approaches have been studied to guide the procurement and design decision-making process. This kind of approach offers risk measurement, rating tools, and compliance checking to identify and rank the vendors by their risk criticality. It is a useful preventive measure that provides a transparent understanding of the security posture in the products, systems, and services of the end-users and helps mitigate the risks prior to the procurement contracts and continuous product development. Cyber resilience complements this measure. It shifts the focus from prevention to recovery by creating a cyber-resilient mechanism to reconfigure the IoT system adaptively to the uncertainties of adversaries and maintain critical functions in the event of successful attacks.  

Many private sectors have for years prioritized efficiency and low cost over security and resilience. In addition, they are agnostic to where these technologies are manufactured and where the associated supply chains and inputs originate. This common practice has resulted in enlarged attack surfaces and many unknown and unidentified threats in the IoT systems. A healthy ecosystem of vendors and suppliers is pivotal to secure and resilient IoT systems. One challenge is that the IoT supply chain is becoming globalized. Manufacturers and material suppliers are geographically diverse, thus increasing the uncertainties and the vulnerabilities of the end-user IoT systems. It is critical to check the compliance of the products from the global supply chain to determine whether they would increase the cyber risk of the IoT users. 

One way to improve the health of the IoT supply chain is to design an IoT system with built-in security and resilience mechanisms. For example, the integration of cyber deception into IoT systems provides a proactive way to detect and respond to advanced and persistent threats. Game-theoretic methods and reinforcement learning techniques have been used to provide a clean-slate approach to designing cyber resilient mechanisms in response to supply-chain attacks. 

\begin{figure}[ht]
\centering
 \includegraphics[width=.8\linewidth]{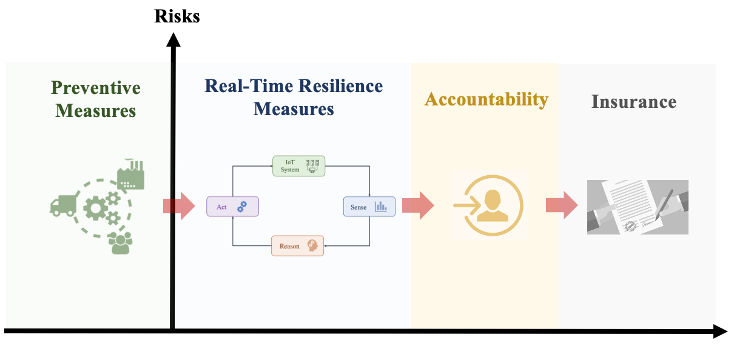}
 \caption{The IoT supply chain can be protected using preventive measures which include compliance checking and auditing. The supply chain resilience can be enhanced by building real-time resilience measures (e.g., detection, adaptation, and reconfigurations). The residual risk as a result of the preventive and real-time resilience measures can be further mitigated by accountability and insurance mechanisms. Accountability is designed to attribute the violations to the suppliers, who will be penalized based on the contract. Insurance is another mechanism to transfer the remaining risks to a third party through an insurance contract. The multi-tier solutions from preventive measures to insurance are interdependent and they create consolidated protection of our IoT supply chain ecosystem. }
\label{fig:4stages}
\end{figure}

Apart from the technological solutions, accountability and cyber insurance are the socio-economic ones that can be used to improve the cyber resilience of IoT end-users. Accountability, in general, is the ability to hold an entity, such as a person or organization, responsible for its actions. An accountable system can identify and punish the party or the system component that violates the policy or the contract. By creating accountable IoT supply chains, we create an ecosystem where each supplier invests in cybersecurity to reduce the cyber risks at each stage of the supply chain. A supplier would be held accountable if the failures of the end-user system are attributed to it. Accountability establishes a set of credible incentives for the suppliers and elicits desirable behaviors that mitigate the cyber risks. Accountability can be viewed as part of the cyber resilience solutions succeeding the technological solutions, especially when the technological resilience measures do not prevent the damages. 

Insurance is another risk management tool to protect the end-users from cyber attacks and failures by transferring their residual risk from an entity to a third party through an insurance contract. It is the last resort when an IoT system cannot be perfectly accountable; i.e., there is inadequate evidence to hold any one of the suppliers accountable, or when the defects in the user's design lead to unanticipated consequences. The residual risks would be evaluated by an underwriter and the coverage can include the losses that arise from ransomware and data theft or incidents caused by failures of IoT devices. Figure \ref{fig:4stages} shows the relationships between preventive cyber measures and resilient cyber measures. The cyber-resilient mechanisms include the technological real-time resilience measures as well as accountability and insurance solutions. They constitute a holistic socio-technical solution to protect the IoT systems from supply-chain threats.

Both accountability and insurance provide an additional layer of protection that reduces the risks of IoT users. Accountability and insurance are system-level issues. We need to take a system-scientific and holistic approach to understand their role in IoT systems and supply chains, which would lead to an integrative socio-technical solution for supply chain security. This chapter provides a quantitative definition to measure and assess the accountability in the IoT supply chain that pertains to the system design, procurement contracts, as well as, vendor description. Despite the focus of the chapter on cybersecurity issues, the definition of accountability can be extended and used for general contexts of supply chain disruptions caused by natural disasters and the defects in the products.

Game theory naturally provides a framework that captures the incentives and penalties through utility functions for multiple interacting agents. In particular, mechanism design theory explicitly provides a quantitative approach to create a reward and penalty mechanism to elicit desirable behaviors at equilibrium. The violations from the desired behaviors would be disincentivized or punished, while the compliance with the rules would be incentivized or rewarded. In this chapter, we leverage these features of game theory to create computational accountability and insurance framework for IoT systems and their supply chain. 

Accountability is a system-level issue that encompasses detection and attribution of the violations or anomalies, multi-agent interactions, asymmetric information, and feedback. Game-theoretic methods provide a baseline for a system-scientific view for accountability. We build a system scientific framework that bridges game theory, feedback system theory, detection theory, and network science to provide a holistic view toward accountability in IoT supply chains. The framework proposed here can be applied to understand accountability in general.

One extension of this chapter is to investigate the concept of collective accountability, where multiple agents are held accountable for the violations. One advantage of such accountability mechanisms is the convenience in identifying the entities to be held accountable and the implementation of the penalties. The disadvantage is that they are not targeted and entities that are not directly linked to the violation of the failures would be also punished.

\section{Literature Review} \label{sec:literature}

Accountability has been studied in many different contexts in computer science \cite{nissenbaum1994computing,feigenbaum2020accountability,feigenbaum2014open}. K\"unnemann et al.  in \cite{kunnemann2019automated} have studied accountability in security protocols. Accountability is defined as the ability of a protocol to point to any party that causes failure with respect to a security property. Zou et al. in \cite{zou2010formal} have proposed a service contract model that formalizes the obligations of service participants in a legal contract using machine-interpretable languages. The formalism enables the checking of obligation fulfillment for each party during service delivery and holds the violating parties for the non-performance of the obligations.  The definition of accountability in these works aligns with the definition in this chapter. An accountable system has the ability to check and verify compliance with the requirements in the agreement and identify the non-conforming behaviors and their parties.

There are several game-theoretic models that are closely related to accountability. For example, inspection games are one class of games where the inspector determines a strategy to examine a set of sampled items from a producer to check whether the producers of the goods violated the standards. The producer aims to set a production strategy to minimize the detection probability while minimizing the cost of maintaining high standards. The inspection games have been used in many contexts such as patrolling, cybersecurity, and auditing. Blocki et al. in \cite{blocki2013audit} have studied a class of audit games in which the defender first chooses a distribution over $n$ targets to audit and the attacker then chooses one of the $n$ targets to attack. It is better for the defender to audit the attacked target than an unattacked target, and it is better for the attacker to attack an unaudited target than an audited one. Rass et al. in \cite{rass2020optimal,rass2016gadapt} have studied a multi-stage cyber inspection game between a network system defender and an advanced persistent threat (APT) attacker. The defender needs to choose an inspection strategy to detect anomalies at different layers of the networks. The attacker's goal is to stay stealthy and find strategies to evade the detection and compromise the target.

Utility-theoretic approaches are useful to capture the incentives of the participants in an agreement and their punishment. In \cite{feigenbaum2014open},  Feigenbaum et al. have formalized the notion of punishment using a utility-theoretic, trace-based
view of system executions. Violation is determined based on the traces of the participants. When there is a violation, the participant is punished. This punishment is captured through a decrease in the utility, relative to the one without the violation. This approach to punishment is often seen in the literature of mechanism design \cite{myerson1981optimal,myerson2008perspectives}. The designer first announces a resource allocation rule and a payment or punishment rule. The participants in the mechanism know the rules and determine the messages that they send to the designer. An incentive-compatible mechanism is one in which the participants will truthfully reveal their private information through the message under the allocation and the punishment rules. In other words, no participants have incentives to lie about their private information under an incentive-compatible mechanism. Mechanism designs have been used in many disciplines to study pricing of resources \cite{farooq2018optimal,zhang2019optimal,zhang2021incentive}, create security protocols \cite{chen2017security}, and design services \cite{zhang2017bi,zhang2019mathtt}.  
The framework that we present in this chapter is built on the mechanism design approach. The utility-theoretic approach conveniently captures the incentives of the suppliers and their behaviors. Furthermore, the mechanism-design approach naturally creates a punishment mechanism to create incentives for truthful behaviors. This type of behavior can be generalized to compliant behaviors in supply chain agreements and contracts.

Our framework builds on this approach and bridges the accountability gap by incorporating the detection mechanism that enables the designer to detect and attribute the non-compliant behaviors. In addition, our framework distinguishes from prior works in accountability by focusing on accountability in system engineering. This problem is instrumental in the development of large-scale IoT systems, where the building blocks of the IoT systems are manufactured or designed by third parties. We integrate the critical component of engineering designs into the accountability problem for IoT systems. The system designs can contribute to accountability. A design is called transparent if it helps identify the cause of the accidents; otherwise, a design makes the accountability inconspicuous. In some cases, the cause of the accidents is not caused by the suppliers but the negligence in the design process. It is important to have a framework that can consolidate multiple factors into the framework and study accountability in a holistic manner. 

\section{Accountability Models in IoT Supply Chain} \label{sec:acc}

\begin{figure}[ht]
\centering
 \includegraphics[width=.8\linewidth]{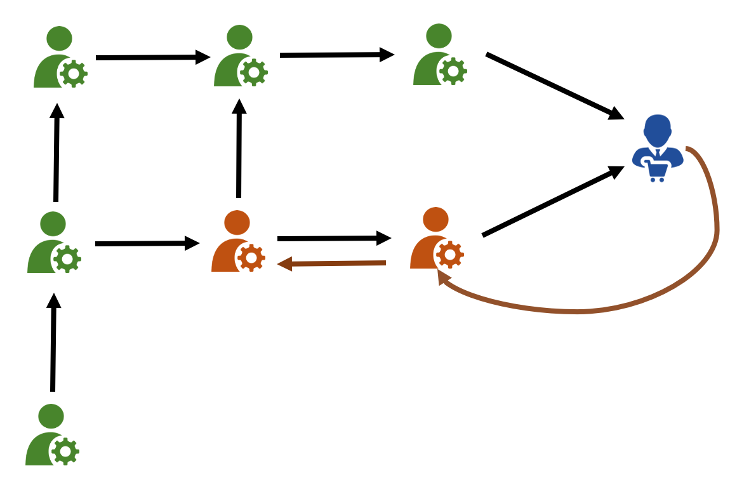}
 \caption{Supply chain accountability: the buyer of the product can identify the supplier of a component who violates the policies or the contracts. The buyer can then use the contract to penalize the identified supplier. The supplier can attribute the violation to his supplier. It is called multi-stage accountability. }
\label{fig:supplychainaccountability}
\end{figure}

\subsection{Running Examples}

We introduce two running examples which will be used in later discussions for illustrations.

{\bf Example I: Uber Autonomous Vehicles}

The Uber incident in Tempe, Arizona is another example of accountability of autonomous vehicles. A pedestrian was struck by an Uber self-driving vehicle with a human safety backup driver in the driving seat. The fatality is caused by the failure of the software system which fails to recognize the pedestrian. Sensor technologies, including radar and LiDAR, are sophisticated enough to recognize objects in the dark. Evidence has shown that the pedestrian was detected 1.3 seconds before the incident and the system determined that emergency braking was required but the emergency braking maneuvers were not enabled when the vehicle is under computer control. The design of the software system is accountable for the death of the pedestrian.  

{\bf Example II: Ransomware attack on smart homes}

A smart home consists of many modern IoT devices, including lighting systems, surveillance cameras, autonomous appliance control systems, and home security systems. The components of each system are supplied by different entities. Smart home technology integrates the components and creates a functioning system that will sense the home environment, make online decisions, and control the system. The camera is accountable if the home security system does not respond to the burglary adequately due to a camera failure. There is an increasing concern about ransomware attacks. Accountability enables the homeowner to mitigate the impact of the ransomware by attributing the attack to a supplier of the IoT devices.

Illustrated in the two examples, IoT supply chain security has a significant impact on the private sector and its customers. Several technologies have been proposed to track the integrity of the supply chain to provide real-time monitoring and alerts of tampering and disruptions. They provide a tool to monitor, trace, and audit the activities of all participants in the supply chain and ensure that the contractually defined Service Level Agreements (SLAs) are followed. The essence of the technologies is to create transparency and situational awareness for the companies. However, the software and hardware tampering is much harder to monitor and track than the physical one. As a result, it creates information asymmetry where the buyers or the systems do not have complete information about their suppliers. As in the Target and the SolarWinds attacks, an attacker can get access to the system through a compromised third-party vendor. It would require proactive security mechanisms to detect and respond to the exploited vulnerabilities. We have seen the emerging applications of cyber deception \cite{pawlick2021game} and moving target defense \cite{jajodia2011moving} in both software and hardware to reduce the information asymmetry and create proactive mechanisms for detection. They are tools that contribute to real-time resilience measures as illustrated in Fig. \ref{fig:accountabilitymodel} and provide inputs for accountability in the next stage.

\subsection{System Modeling}

\begin{figure}[ht]
\centering
 \includegraphics[width=.8\linewidth]{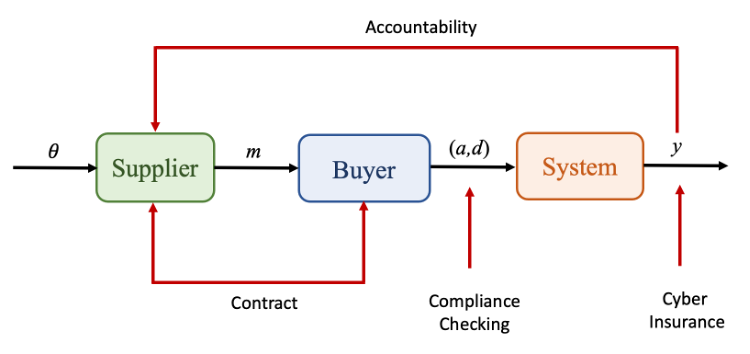}
 \caption{A supplier of type $\theta$ provides a description $m$ of the product to a buyer who will make a procurement decision $a$. The system designer develops a design $d$ to integrate all the components to form a functioning system. The system, as a result, yields an observable performance $y$. The supply chain is said to be accountable if the malfunction of the system can be attributed to the supplier who has misled the system designer. The supply chain risk can be mitigated at three stages. The first stage is compliance checking before the procurement. The buyer can check whether the description of the product complies with the standards, regulations, and requirements. The second stage is the contracting stage. The buyer can make a contract that specifies the penalty or the consequences if the supplier does not fully disclose the product information. It will allow the buyer to hold the supplier accountable when the root cause of the malfunction is at the supplier. The third stage is cyber insurance. The buyer can purchase cyber insurance to mitigate the financial impact of the malfunction. The financial risk is partially transferred to the insurer. }
\label{fig:accountabilitymodel}
\end{figure}

In this section, we provide a stylized model and a quantitative approach to accountability. Fig. \ref{fig:accountabilitymodel} describes three stages of interactions. At the first stage, a supplier interacts with a buyer to agree on an SLA contract. The supplier is characterized by the private information $\theta\in\Theta$, which is a true description of the product of the supplier. For example, the supplier is aware of the true security level and investment in the product but may not disclose the information to the buyer. The supplier sends the buyer a message $m\in M$, which is the informed description of the product. The description can prevaricate, hide, or sometimes lie about the security information that would be useful in the procurement decisions. We say that the supplier truthfully reports the product when $\theta = m$; otherwise, we say that the supplier misinforms the buyer. This misinformation can be unintended or intentional. In the case of intentional behaviors, the supplier sends a manipulative message when he knows his true type. For example, some foreign suppliers do not fully disclose the information of their product with the aim to attract US customers due to its low cost. In the case of unintended behaviors, the supplier may not be aware of the vulnerabilities of the product and sends a description based on his perceived information. In this case, we can assume that the private information $\theta$ is a function $\rho: \Theta\times \mathcal{W} \mapsto\Theta$ of the truth and uncertainties, i.e., $\theta= \rho(\theta_{\textrm{t}}, w_{\textrm{t}})$, where $\theta_{\textrm{t}}\in \Theta$ is the true value unobservable by the supplier and $w_{\textrm{t}}\in \mathcal{W}$ is the bias, modeled as a random variable, unknown to the supplier. This bias can be interpreted as the uncertainties introduced by nature or a stealthy attacker that has unknowingly changed the security attributes of the product. In both cases of unintended and intentional behaviors, it is sufficient to assume that the type known to the supplier is $\theta$.

Based on the product description $m$, the buyer can make purchase decisions. Let $a = 1$ denote the decision of adopting the product of the vendor and $a = 0$ otherwise. The decision rule $\alpha: \mathcal{M} \mapsto [0,1]$ yields the probability of purchase based on the received description, i.e. $\alpha(m)=\text{Pr}(a=1|m)$. This can be interpreted as the purchase preference from historical records. If the buyer decides to adopt the product, then he determines how the product is designed and integrated into the system. Here, we assume that the user and the designer belong to the same organization and hence the procurement and design decisions are made jointly. In other words, the user and the designer can be viewed as the same decision entity who coordinates the design and procurement. In practice, the engineers design the systems and send the procurement department the specifications and requirements for the needed materials and components.

An IoT system consists of many components. We can classify the components into five major categories: sensing, computation, control, communications, and hardware. The sensing component allows the system to provide information about the environment, for example, the LiDAR and temperature sensors. The computation units provide functions and services for information processing and computations, for example, cloud services and GPUs. The control components are used to instrument and actuate the physical systems, for example, temperature adjustment and remote control. Communications provide the information and data transmission among IoT components, e.g., LoRa and ZigBee wireless communications. The hardware refers to the physical systems that underlie the IoT network, for example, the manufacturing plant and the robots.

The designer builds an IoT system using a blueprint $\delta: \mathcal{M} \mapsto \mathcal{D}$, which yields a design $d=\delta(m), d \in \mathcal{D}$ based on the device descriptions and specifications provided by the supplier. The system design leads to a performance $y\in \mathcal{Y}$. For example, in Example I, the designer develops a software system that integrates sensors, control algorithms, and the car. Safety is a critical performance measure of autonomous vehicles. It can be measured by the rate of accidents experienced by vehicles as of now. Here, we model the performance as a random variable. Given $\alpha$ and $\delta$, the distribution of the performance random variable is $p_y(y;\theta,\alpha(m),\delta(m))$, $p_y:\Theta\times \mathcal{M}\mapsto \Delta\mathcal{Y} $. Using Bayes' rule, we arrive at
\begin{equation}
p_y(y; \theta, \alpha(m), \delta(m)) = p^\theta_y(y;\alpha(m),\delta(m)|\theta)p_\theta(\theta),
\end{equation}
where $p_\theta(\cdot)\in\Delta\Theta$ is the prior distribution of the type of product; $p^\theta_y(y; \alpha(m),\delta(m)|\theta)$, $p^\theta_y:\mathcal{M}\mapsto\Delta\mathcal{Y}$, is an indication of all possible system performances given the attribute of product $\theta$. Note that the performance implicitly depends on $m$. The true performance of the system is determined by the true attribute of the product and the procurement and design decisions, which are made based on $m$. We denote $p_I=p_y(y; \theta, \alpha(\theta), \delta(\theta))$ as the ideal system performance when the design and procurement decisions are made given a truthful supplier, i.e., $m=\theta$.

Without knowing the true attributes of the product $\theta$, the performance anticipated by the buyer is denoted by $q_y=p_y(y; m, \alpha(m), \delta(m))$. When $m\neq \theta$, there is a difference between the observed performance $p_y$ and the anticipated one $q_y$. The buyer can perform hypothesis testing based on the sequence of observations $y_1, y_2, \cdots,$ by setting up $H_0$ as the hypothesis that the observations follow the distribution $q_y$ and $H_1$ otherwise. For example, in Example I, this decision is particularly important when $y_i$ represents malfunctions or accidents for each trial test driving. If the malfunction is not expected by the designer, then there is a need to find out which supplier is accountable for the accidents or, in the case of a single supplier, whether the supplier should be held accountable.

\subsection{Accountability Investigation}

One critical step of accountability is the ability to attribute the performance outcomes to the supplier. We start with the accountability of a single supplier with binary type $\Theta = \{0,1\}$ and assume the message space is the same as type space $\mathcal{M} = \Theta$. Consider a sequence of repeated but independent observations $Y^k=\{y_1, y_2, \cdots, y_k\}$, $k\in \mathbb{N}$. A binary accountability investigation is performed based on $Y^k$. Based on the received $m$, hypothesis $H_0$ is set to be the case when the observations follow the anticipated distribution $q_y$ and $H_1$ otherwise. Depending on whether $H_0$ or $H_1$ holds, each observation $y_i$ admits the following distribution

    \begin{align}
            &H_0: \, y_i \sim f_m(y|H_0) = p_y(y;m,\alpha(m),\delta(m)), \\
            &H_1: \, y_i \sim f_m(y|H_1) = p_y(y;\neg m,\alpha(m),\delta(m)).
    \end{align}

The optimum Bayesian investigation rule is based on the likelihood ratio, which is denoted by
    \begin{align}
         L(Y^k)=\prod_{j=1}^k\frac{p_y(y_j;\delta(m)|\neg m)p_\theta(\neg m)}{p_y(y_j;\delta(m)|m)p_\theta(m)},
     \end{align}
where we omit the purchase decision because the performance can only be observed when $a=1$ and $\alpha(m)=Pr[a=1|m]$ is the same under both hypotheses. The likelihood ratio test (LRT) provides the decision rule that $H_1$ is established when $ L(Y^k)$ exceeds a defined threshold value $\tau_k\in\mathbb{R}$; otherwise, $H_0$ is established. It can be formulated by the equation
\begin{align}
    L(Y^k)  \mathop{\gtreqless}_{H_0}^{H_1} \tau_k.
    \label{LRT}
\end{align}

One critical component in accountability investigation is the prior distribution over hypotheses, which indicates the reputation of the supplier. Without knowing the true distribution of the type, we argue that reputation is sufficient knowledge to determine the accountability of the supplier. Here we give the definition of reputation over a binary type space, but the definition can be extended to multiple type space accordingly.

\begin{definition}[Reputation]
\label{reputation}
The reputation of the supplier $\pi\in\Delta \mathcal{H}$ is a prior distribution over all hypotheses. In binary case, $\pi_0 = \text{Pr}[H_0]$ is the prior probability that the supplier truthfully report and $\pi_1=\text{Pr}[H_1]$ otherwise, with  $\pi_0+\pi_1=1$.
\end{definition}

Assume the cost of the investigation is symmetric and incurred only when an error occurs. In the binary case, the optimum decision rule will consequently minimize the error probability, and the threshold value $\tau_k$ in LRT will reduce to 
\begin{align}
    \tau_k = \pi_0/\pi_1.
\end{align}

\begin{definition}[Accountability] \hfill
\label{accountability}
\begin{enumerate}
\item 
Given an investigation rule,  i.e., the threshold $\tau_k$, the accountability $P_A \in [0,1]$ is defined as the probability of correct establishment of hypothesis $H_1$ based on the observations $Y^k$ and message $m$, which is given by
\begin{align}
    P_A(\tau_k)= \int_{\mathcal{Y}_1} f_m(Y^k|H_1)dy^k,
\end{align}
where $\mathcal{Y}_1$ is the observation space where $\mathcal{Y}_1 = \{Y^k: L(Y^k)\geq \tau_k\}$.

\item The wronged accountability $P_U\in[0,1]$ is defined as the probability of a false alarm that $H_1$ is established while the underlying truth is $H_0$. Consider the threshold $\tau_k$ and observations $Y^k$, $P_U$ is given by
\begin{align}
    P_U(\tau_k)= \int_{\mathcal{Y}_1} f_m(Y^k|H_0)dy^k.
\end{align}
\end{enumerate}
\label{def:1}
\end{definition}

We call a supplier $\eta$-unaccountable if $P_A\leq \eta$, for a threshold accountability $\eta\in[0,1]$ chosen by the investigator. In this case, the system does not have strong confidence that the observed accidents are caused by the supplier. We call a system $\epsilon$-nontransparent if $P_A\leq \epsilon$, for a given small $\epsilon \in [0,1]$. That is, the system is close to being unable to hold the vendor accountable for the accidents. 

The performance of the accountability investigation will be evaluated in terms of $P_A$ and $P_U$. Ideally, we would like to conduct error-free accountability testing where $P_A$ is close to one and $P_U$ is close to zero (correctly identify accountable supplier without making mistake). However, the definition above leads to a fundamental limit on the accountability of the supplier. Except for situations where the observations $Y^k$ under $H_0$ and $H_1$ are completely separable or the number of observations $k$ goes to infinity, the performance of the accountability testing will be restricted within a feasible region.

    \begin{figure}[ht]
            \centering
            \includegraphics[width=0.9\textwidth]{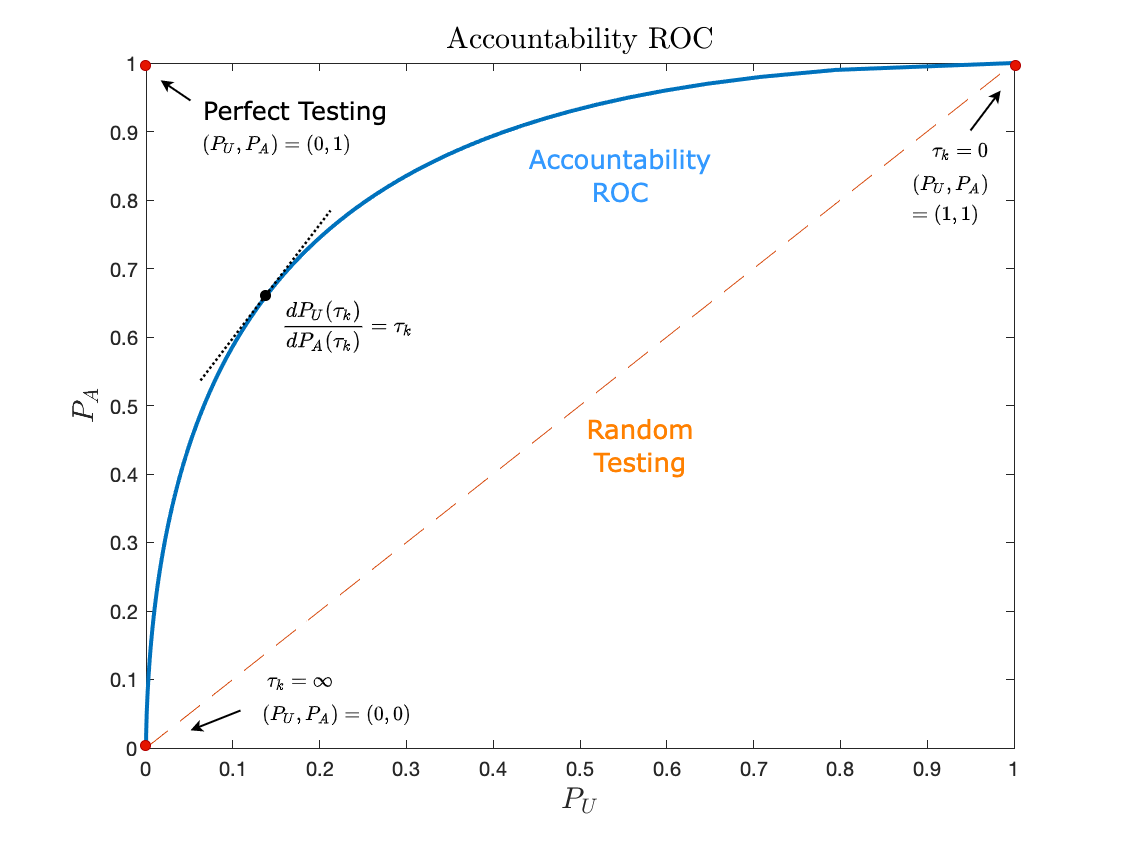}
            \caption{Accountability receiver operating characteristics (AROC).}
            \label{fig:aroc}
    \end{figure}

\begin{definition}[Accountability Receiver Operating Characteristic]
Accountability\\Receiver Operating Characteristic (AROC) is a plot which describes the relationship between achievable accountability $P_A$ and wronged accountability $P_U$ in the square $[0,1]\times[0,1]$.
\end{definition}

As shown in Fig.~\ref{fig:aroc}, if we conduct LRT in accountability investigation, the AROC curve depicts the testing performance with respect to different threshold values $\tau_k$. Similar to traditional binary hypothesis testing, the AROC curve under proper design preserves the following properties \cite{levy2008binary}.
\begin{property}[AROC]
AROC curve under proper design has the following properties:
\begin{enumerate}
    \item[(1)] $(P_U,P_A)=(0,0)$ and $(1,1)$ belong to the AROC. 
    \item[(2)] The slope of the AROC curve $dP_A(\tau_k)/dP_U(\tau_k)$ is equal to the threshold $\tau_k$.
    \item[(3)] The AROC curve is concave and the feasible domain of $(P_U,P_A)$ is convex.
    \item[(4)] $P_A(\tau_k)\geq P_U(\tau_k)$, $\forall \tau_k\in[0,+\infty)$.
\end{enumerate}
\end{property}

\begin{remark}
The likelihood ratio lies in the region between zero and infinity. 
If we set the threshold $\tau_k$ in LRT to zero, investigator will classify any performance results into hypothesis $H_1$ (misinformation). Both accountability $P_A$ and wronged accountability $P_U$ will approach to one, as $(P_U,P_A)=(1,1)$. Similarly, if we set $\tau_k$ in LRT to infinity, investigator will classify any performance into hypothesis $H_0$ (truthfully report), resulting in $(P_U,P_A)=(0,0)$.
\end{remark}
\begin{remark}
Property (3) and (4) are satisfied under the proper design, i.e. the test is ``good'' with $P_A\geq P_U$. For a ``bad'' test with $P_A< P_U$, because of the real meaning behind the hypothesis, we cannot simply reverse the performance distribution as in traditional hypothesis testing. Instead, we need to re-construct the investigation and find another performance metric that can properly distinguish the misinformation between the supplier and buyer.
\end{remark}

It is worth noting that as the threshold $\tau_k$ increases, the accountability of the supplier $P_A$ increases. However, according to the aforementioned properties, it would also increase wronged accountability $P_U$ when the accidents are not caused by the vendor. There is a fundamental trade-off between accountability $P_A$ and wronged accountability $P_U$ depending on the accountability investigation. One way to evaluate the investigation performance is the area under the AROC curve (AUC). AUC is a measure of investigation capability \cite{wickens2001elementary}, which provides a simple figure of merit to represent the degree of separability between two hypotheses. 
\begin{align}
    AUC(\tau_k) = \int_0^1 P_A(\tau_k) \,\, dP_U(\tau_k)
\end{align}
This value varies from 0.5 to 1. When AUC equals $0.5$, the designed investigation has no separation capability, which means the performance of the test is no better than flipping a coin. This is corresponding to the case when $P_A(\tau_k)=P_U(\tau_k)$ for all possible threshold $\tau_k$. Ideally, an excellent test will produce an AUC equal to one. In this situation, the accountability investigation can completely distinguish between two hypotheses, thus correctly identifying the supplier who should be accountable for the accidents.

Unfortunately, in realistic investigation tasks, it is hard to obtain the exact computation of AUC. Analyzing the upper and lower bounds of AUC will help the investigator to describe the performance of the designed test. Shapiro in \cite{shapiro1999bounds} provides an upper bound and lower bound on binary testing. Consider equally likely hypotheses with $\tau=1$, the probability of error $P_e\in[0,1]$ is defined as 
\begin{align}
    P_e = \frac{P_U(\tau=1)}{2} + \frac{1-P_A(\tau=1)}{2}.
\end{align}
Due to the convexity of the AROC curve, the bounds of the AUC can be described as
\begin{align}
    1-P_e \leq AUC \leq 1-2P_e^2.
\end{align}

\subsection{Model Extensions} 
This framework can be extended to multiple product types and multiple suppliers. The accountability needs to point to any suppliers that cause failures under the hypothesis. In this section, we provide several testing frameworks and the definition of accountability accordingly.

\subsubsection{Single Supplier with Multiple Types}

Consider the product from the supplier with $T\in\mathbb{N}$ possible types, $\Theta = \{\theta_1,\theta_2,\dots,\theta_T\}$. Based on the received message $m=\theta_m$, hypotheses $\{H_1,H_2,\dots,H_T\}$ can be constructed by the investigator such that the performance observation $y$ under each hypothesis $H_t$ admits
    \begin{align}
            &H_t: \, y \sim f_m(y|H_t) = p_y(y;\theta_t,\alpha(\theta_m),\delta(\theta_m)),
    \end{align}
for $1\leq t\leq T$. The distribution under hypothesis $H_t$ describes the system performance if the buyer makes purchase and designs based on the message $\theta_m$ while the underlying true product type is $\theta_t$. In this case, the only anticipated performance by the buyer follows $H_m$. Any other observation distribution $H_{t\neq m}$ will attribute to the accountability of the supplier. Investigation could be conducted through M-ary hypothesis testing. For a single supplier with multiple product types, we can define the accountability as follows.

\begin{definition}[Accountability with multiple types]
Given a detection rule $\lambda$, the received message $m$ and observations $Y^k$,  the accountability  for a single supplier with multiple product types is defined as
\begin{align}
    P_A(\lambda)= \sum_{t\neq m, 1\leq t\leq T} \int_{\mathcal{Y}_t} f_m(Y^k|H_t)dy^k,
\end{align}
where $\mathcal{Y}_t$ is the observation space we classify the observations as $H_t$.
\end{definition}

If we assume the investigation cost is symmetric and only occurs with error, this gives a MAP decision rule and the performance of the accountability testing can be evaluated through the error probability as
\begin{align}
    P_e = \sum_{1\leq t\leq T} \text{Pr}(E|H_t)\pi(t),
\end{align}
where $E$ denote the error event and $\pi(\cdot)\in\Delta\Theta$ is the prior probability that $H_t$ will happen,  which represents the reputation of the supplier.

\subsubsection{Multiple Suppliers}
\label{sec:multiple}

\begin{figure}[ht]
\centering
 \includegraphics[width=.8\linewidth]{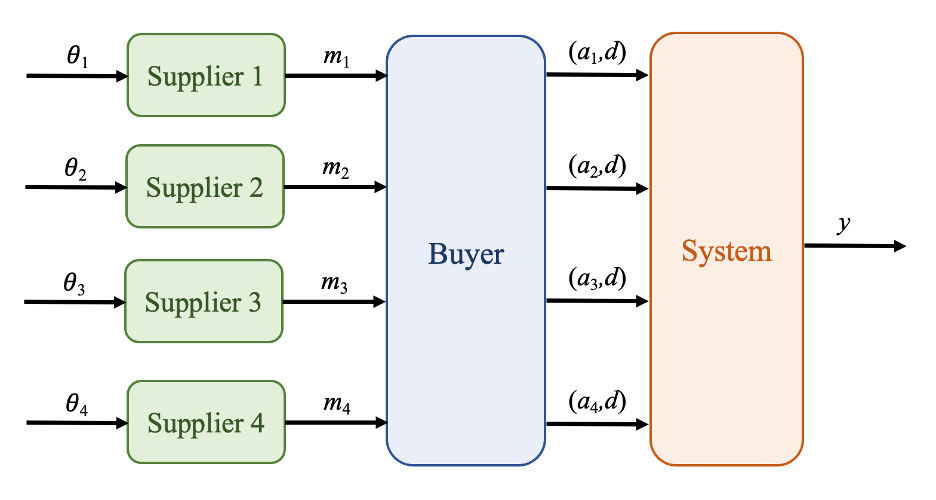}
 \caption{Extension of the model to multiple suppliers. }
\label{fig:extension}
\end{figure}

In IoT system design with multiple suppliers, accountability testing needs to point to any suppliers that cause failures under the hypothesis. To simplify the illustration, we consider the case where the component from each supplier may have binary types $\theta_i\in\{0,1\},\forall i\in\mathcal{I}$. Consider the problem with $N$ vendors in the supply chain. Each supplier $i\in \mathcal{I} = \{1,2,\dots,N\}$ with true product type $\theta_i$ will send a message $m_i\in\mathcal{M}_i$ to the buyer to make purchase decision $a_i\in\{0,1\}$ and determine the overall design $d\in\mathcal{D}$. The process is illustrated in Fig.~\ref{fig:extension}. We can construct hypotheses as a vector 
\begin{align}
    H_j = (h_1,h_2,\dots,h_N), \quad h_i = \mathds{1}(m_i\neq \theta_i)\,  \forall i\in\mathcal{I},
\end{align}
where each element $h_i$ is an indicator of whether supplier $i$ truthfully reports or not, and the subscript $0\leq j\leq 2^N-1$ is the decimal number of the binary combination in the vector. The hypothesis vector indicates which supplier(s) should be accountable for the accident. When the performance distribution under each hypothesis is distinguishable, the investigation could be conducted through M-ary hypothesis testing. Otherwise, we can consider decentralized investigation as described in the following.

\begin{figure}[ht]
\centering
 \includegraphics[width=.75\linewidth]{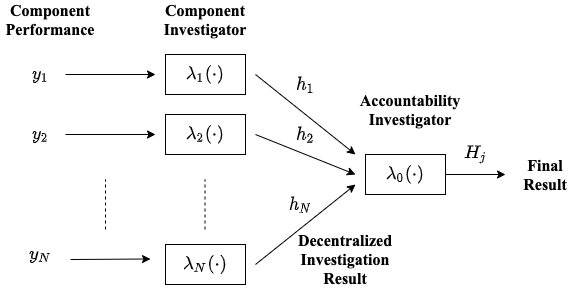}
 \caption{Decentralized testing  }
\label{fig:decentralized}
\end{figure}

Consider a decentralized accountability investigation with $2^N$ hypothesis $H_0,..,H_{2^N-1}$ and prior reputation $\pi(H_0),\dots, \pi(H_{2^N-1})$, respectively. Suppose we have $N$ suppliers providing components to the system. Each component investigator $\lambda_i$ is inspecting the performance related to the product provided from the vendor $i$. 
In practice, we can design the independent tests for each component to determine the accountability of supplier $i$. We can control the other parts $(j\neq i)$ to be known and fixed products in test design and focus on the binary hypothesis testing with respect to component $i$.

Each component investigator receives observations $y_i$, which is a random variable taking values in a set $\mathcal{Y}_i$. The local investigator will conduct accountability testing through $\lambda_i:\mathcal{Y}_i\mapsto \{0,1\}$ and output a binary decision variable $h_i=\lambda_i(y_i)$, which indicates whether supplier $i$ should hold accountable for the accident. This reduces the problem to $N$ parallel binary hypothesis testing with each supplier, and the accountability of each supplier then will be the same as we defined in \ref{def:1}. The final investigator determines which hypothesis will be established based on received information, $\lambda_0:\{0,1\}^N \to \{0,1,\dots,2^N-1\}$. It has been shown in \cite{tsitsiklis1989decentralized} and \cite{nguyen2008distributed} that there exists an optimal detection rule if each testing observations are independent or conditionally correlated under each hypothesis.

\section{Case Study 1: Autonomous Truck Platooning}
\label{sec:aeb}

In the following section, we will provide a detailed case study in autonomous truck platooning with adaptive cruise control (ACC) system. This example illustrates the case when the true performance is unknown to the investigator. We will discuss the accountability of the ranging sensor supplier in the case of a collision. 

\subsection{Background}

With the rapid development of autonomous vehicles, safety is one of the main priorities for manufacturers. As estimated by the World Health Organization (WHO), the number of annual road deaths with collision has reached 1.35 million worldwide \cite{world2018global}. The recent incident in Tempe, Arizona, has thrown a spotlight on the safety of autonomous vehicles. The Uber self-driving test car caused the death of the pedestrian because of the failure of braking control by the autonomous driving system. The investigation of accountability is crucial to determine the cause of the collision and provides insights for future car design. 

In this case study, we consider the task of autonomous truck platooning with Adaptive cruise control (ACC) system. Adaptive cruise control is a driver assistance technology that maintains a safe following distance between the vehicle and traffic ahead without any intervention by the driver. If the preceding truck is detected traveling too slowly or too close, the ACC system will react by automatically activating the brakes and mitigating potential collisions. Brake control is determined based on the relative distance, relative velocity, and the acceleration of leading and the following truck. The speed and acceleration of both vehicles can be measured by built-in speed sensors and accelerometers. Ranging sensors, including radar and LiDAR, are used for distance detection in the ACC system. The upper-level control system uses the measurements of the sensors to interpret the driving environment, and trigger appropriate brake action to mitigate collision \cite{stockle2018robust}. Thus, the detection range and precision of the ranging sensor are critical in ACC design. Defective ranging sensors could cause severe consequences and should be held accountable in case of such a collision.

\subsection{Vehicle Dynamics Model}

\begin{figure}[ht]
            \centering
            \includegraphics[width=0.6\textwidth]{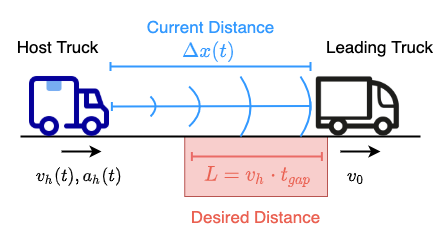}
            \caption{Host truck with ACC system following the leading truck.}
            \label{fig:1}
        \end{figure}
        
To illustrate the accountability of the ranging sensor in this framework, we first introduce the dynamics model of the problem. Consider the testing scenario in Fig.~\ref{fig:1}, where the host truck equipped with ACC system is approaching the preceding vehicle. The control goal of the ACC system is to maintain the desired safe distance from the leading vehicle. The desired distance $L$ is normally determined by \textit{Constant time gap} spacing policy in ACC systems, which guarantees the individual vehicle stability and string stability \cite{stockle2018robust}. 
\begin{align}
    L = v_h \cdot t_{gap},
\end{align}
where $v_h$ is the speed of the host vehicle and $t_{gap}$ is the constant desired time gap.

Denote $x_i, v_i, a_i$ as the position, velocity and acceleration of the leading ($i=l$) or host ($i=h$) vehicle. We assume the leading vehicle is at constant speed $v_l(t) = v_0$. The system state vector $\mathbf{x}(t)$ and control vector $\mathbf{u}(t)$ are defined as follows \cite{wang2014rolling}.
\begin{align}
    \mathbf{x}(t) = \begin{bmatrix} \Delta x(t) - L, &  \Delta v(t)\end{bmatrix}^T,
    \quad 
    \mathbf{u}(t) = \begin{bmatrix} a_h(t)\end{bmatrix},
\end{align}
where $\Delta x(t) = x_l(t) - x_h(t)$ is the current distance and $\Delta v(t) =  v_l(t) - v_h(t)$ is the relative speed between the leading and following vehicles. The state space representation of the system can be written as 
\begin{align}
    \dot{\mathbf{x}}(t) &= A\mathbf{x}(t)+B\mathbf{u}(t),\\
    y(t) &= C\mathbf{x}(t)+w(t),\label{eq:state}
\end{align} The matrices are given by
\begin{align}
    A = \begin{bmatrix}
    0 & 1\\
    0 & 0 
    \end{bmatrix},
    \quad 
    B = \begin{bmatrix}
    -t_{gap} \\
    -1 
    \end{bmatrix},
    \quad 
    C = \begin{bmatrix}
    1 & 0
    \end{bmatrix},
\end{align}
where $y(t)= \Delta x(t) - L+w(t)$ is the noisy control error between the desired distance and current distance; $w(t)$ is the observation noise. We assume the observation disturbance is modeled by an additive white Gaussian noise,
\begin{align}
    w(t) = \mathcal{N}(0,\sigma^2).
    \label{eq:r}
\end{align}
The variance $\sigma^2$ indicates the influence of the measurement environment. 
The intuition behind using the Gaussian noise model is that it gives a good approximation of the natural processes. If a specific distribution of measurement error is given, the noise model can be changed accordingly and the accountability testing framework will still work.

The optimal control can be achieved through linear quadratic regulator (LQR) control. We define the cost function with zero terminal cost as
\begin{align}
    J = \frac{1}{2}\int_{t=0}^{\infty} \mathbf{x}(t)^TQ \mathbf{x}(t) + \mathbf{u}(t)^T R \mathbf{u}(t) \, dt,
\end{align}
where the diagonal weights
\begin{align}
    Q = \begin{bmatrix}
    w_1 & 0\\
    0 & w_2 
    \end{bmatrix},
    \quad 
    R = \begin{bmatrix}
    1
    \end{bmatrix}.
\end{align}
The goal of the controller is to regulate the state towards $(0,0)^T$. 
The feedback optimal control low is given as
\begin{align}
    u(t) = -R^{-1}B^T P\mathbf{x}(t)
\end{align}
where $P$ is the solution to the following associated algebraic Riccati equation:
\begin{align}
    0 = PA+A^T P+Q - PBR^{-1}B^T P.
\end{align}

The aforementioned vehicle dynamics model and optimal control describe the system design $\delta$ of the final ACC system based on the information provided by the supplier. Different control methods and system design can be implemented to achieve the same goal. In the following section, we assume that this system design is not the cause of the collision and purely focuses on the accountability of the sensor supplier. 

\subsection{Accountability Testing}\label{sec:acc_test}
The true product attributes play an important role in control system design. From the previous section, the optimal control of the system depends on the correct distance detection between the two objectives. Thus, the sensor with degraded detection result should hold accountable if the ACC system fails to maintain the safety distance and causes a collision. To attribute the ACC system performance to the ranging sensor supplier, we conduct the following accountability testing with respect to the ranging sensor.

For the simplicity of the model, we consider two types of ranging sensor $\theta\in\Theta = \{0,1\}$, which differ in the detection precision. We assume the sensor with type $\theta = 1$ is functioning normally, as the detection result $r_1(t) = \Delta x(t)$; while the sensor with type $\theta=0$ is malfunctioning with detection result $r_0(t) = \Delta x(t)+e_d$. The value $e_d$ is the detection error of the ranging sensor. The damaged sensor will put the host vehicle at risk of collision, since the actual distance is closer to the detection result. 

The true property of the sensor is private information to the supplier, which is not revealed to the system designer. The supplier should hold accountable for a collision if there exists misinformation between the product description $m$ and true product property $\theta$. Note that the misinformation can be unintended or intentional. We would like to determine whether the ranging sensor supplier should hold accountable for such an accident.

        \begin{figure}[ht]
            \centering
            \includegraphics[width=0.7\textwidth]{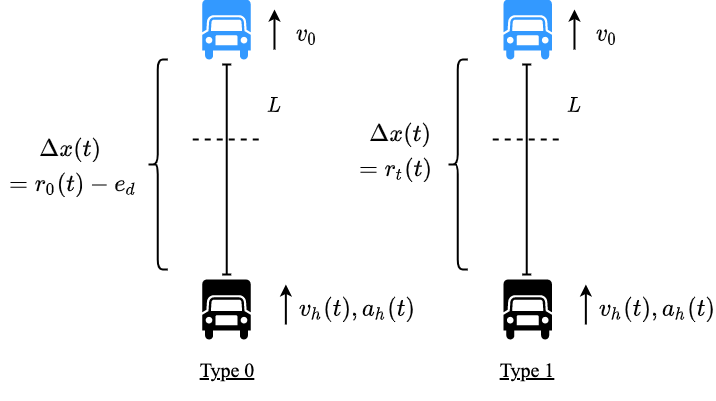}
            \caption{Accountability testing with different sensor types.}
            \label{fig:2}
        \end{figure}

Consider the testing scenario in Fig.~\ref{fig:2}. The distance detection result from the sensor will be the input of the state vector as
\begin{align}
    \mathbf{x}(t) = \begin{bmatrix} r_\theta(t) - L, &  \Delta v(t)\end{bmatrix}^T.
\end{align}
We use the final distance control error as the performance $y$ of the ACC system when testing. Suppose the supplier report $m=1$ when signing the contract. Consider a noisy observation results $y$ as described in (\ref{eq:state}), then the performance should follow
\begin{align*}
    y\sim p_y(y;1,\alpha(1),\delta(1)) = N(0,\sigma^2).
\end{align*} 
This is the anticipated distribution of the observations if the supplier truthfully report the product type ($m = \theta = 1$). On the other hand, if the supplier misinforms the buyer ($m \neq \theta=0$), the performance should follow 
\begin{align*}
    y\sim p_y(y;0,\alpha(1),\delta(1)) = N(- e_d,\sigma^2) 
\end{align*}
The negative distance control error suggests that the distance between two vehicles is smaller than the desired safety distance requirement $L$, which can lead to a potential collision.

We set up the following hypotheses to estimate the accountability of the supplier who reports $m=1$. Let $\mathbf{Y}=[y_1,y_2,\dots,y_N]\in \mathbb{R}^N$ be a vector of independent identically distributed observations $y_k$ $(1\leq k\leq N)$ of the aforementioned testing scenarios. 
    \begin{align*}
            &H_0: \, \mathbf{Y} \sim N(-e_d,\sigma^2\mathbf{I}_N) \\
            &H_1: \, \mathbf{Y} \sim N(0,\sigma^2\mathbf{I}_N)
        \end{align*}
where $\mathbf{I}_N$ is the identity matrix of size $N$. To keep the consistency with other studies, $H_1$ represents the case that the supplier truthfully report. $H_0$ suggests there exists misinformation between the reported product description $m$ and true product type $\theta$. The supplier will be accountable if the investigator correct detected that hypothesis $H_0$ should be established.

Assume the cost of the decision is symmetric and incurred only when an error occurs. The reputation of the supplier follows $[\pi_0, \pi_1]$. In Bayesian binary hypothesis testing, LRT will compare the likelihood ratio to threshold $\tau=\pi_0/\pi_1$. The result suggests that the hypothesis $H_0$ will be established if the sample mean $S$ is smaller than the testing threshold $\eta$, as shown in the following
    \begin{align}
            S=\frac{1}{N}\sum_{i=1}^N y_i  \mathop{\gtreqless}_{H_0}^{H_1} \eta
            \label{eq:s}
        \end{align}
where
        \begin{align}
            \eta = \frac{e_d}{2}+\frac{\sigma^2 \ln(\tau)}{N e_d}
            \label{eq:eta}
        \end{align}
Given the decision rule and supplier's reputation ratio $\tau$, the accountability and wronged accountability of the sensor supplier who reported $m=1$ is 
        \begin{align}
            P_A(\tau) &= \int_{\mathcal{Y}_0} f_1(y|H_0)dy = 1-Q\left(\frac{d}{2}+\frac{\ln(\tau)}{d}\right) 
            \label{eq:pa}\\
            P_U(\tau) &= \int_{\mathcal{Y}_0} f_1(y|H_1)dy = Q\left(\frac{d}{2}-\frac{\ln(\tau)}{d}\right)
            \label{eq:pu}
        \end{align}
where $Q(x)$ is the Gaussian Q function and  $d=N^{1/2}e_d/\sigma$ \cite{levy2008binary}.

\subsection{Parameter Analysis}   
The accountability of the sensor supplier helps the investigator to determine whether the failure of the ACC system should be attributed to the sensor. Since the accountability depends on parameters such as sampling size $N$, environmental observation noise variance $\sigma^2$ and sensor range difference $e_d$. In this section, we discussed several numerical results under different cases.

\begin{figure}[ht]
\centering     
\subfigure[Accountability $P_A$]{\label{fig:diffpa}\includegraphics[width=0.48\textwidth]{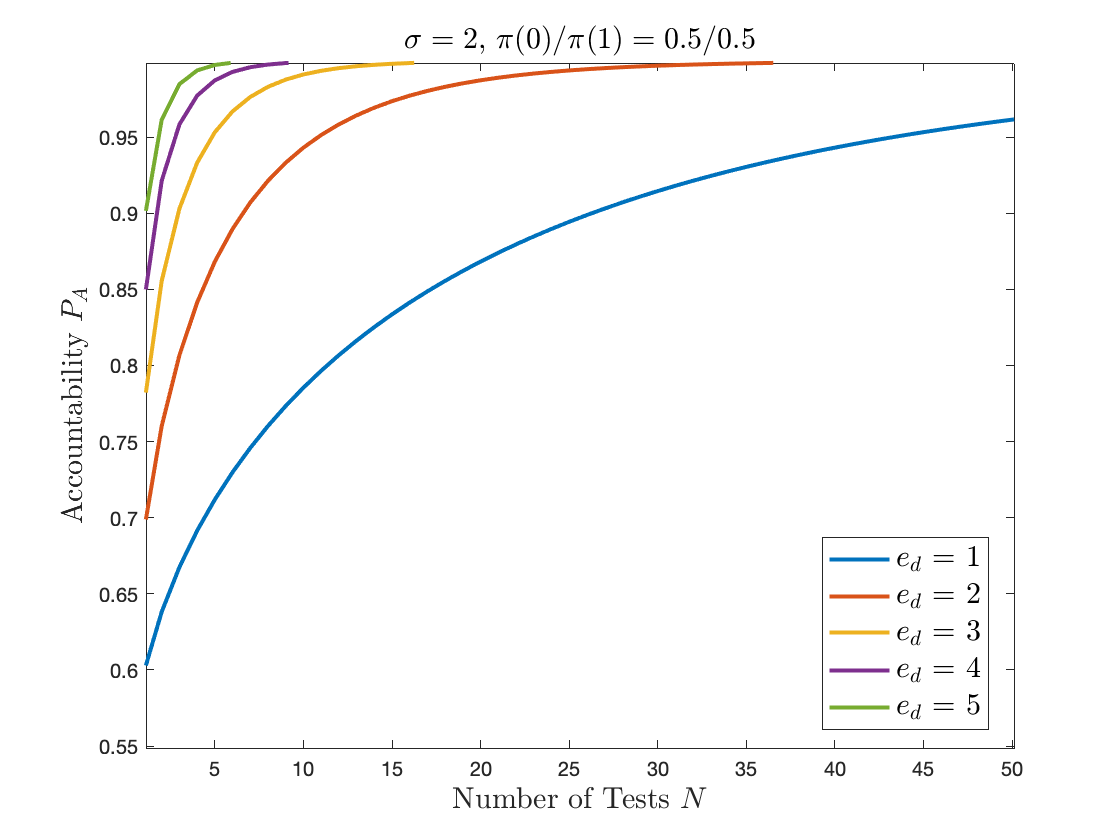}}
\subfigure[Wronged Accountability $P_U$]{\label{fig:diffpu}\includegraphics[width=0.48\textwidth]{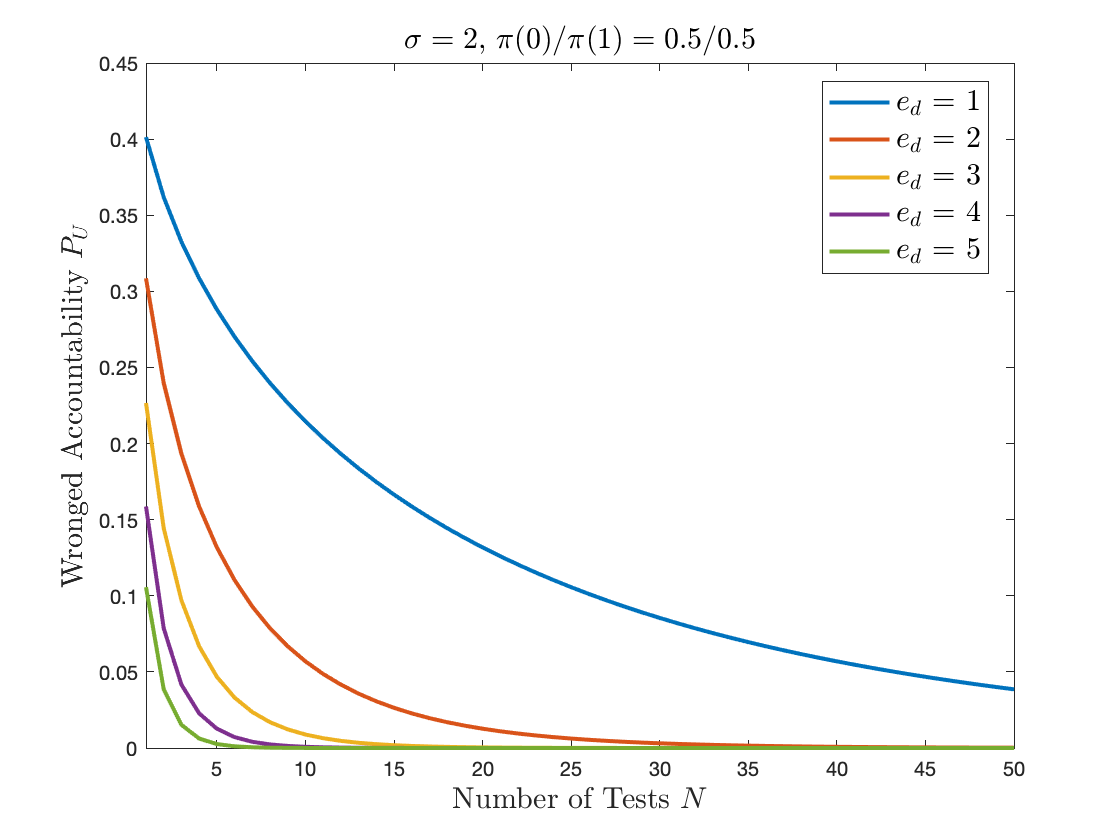}}
\caption{Different sensor range difference ($\sigma=2$, $\pi_0/\pi_1=0.5/0.5$)}
\label{fig:diff}
\end{figure}

Figure~\ref{fig:diff} depicts the influence of the number of tests $N$ and sensor detection error $e_d$ on the accountability. First, we notice that the $P_A\to 1$ and $P_U\to 0$ as the number of tests $N$ increases. This phenomenon indicates more testing will produce a more accurate detection of the supplier's accountability. From equation (\ref{eq:s}), we note that the observation means $S$ converges almost surely to the expected mean of each hypothesis as $N\to\infty$. Besides, the second term in the testing threshold $\eta$ vanishes, and we end up comparing the expected mean of $Y$ to the middle point $e_d/2$ of two hypothesis means. 

The influences of sensor detection error  $e_d$ is also illustrated in Fig~\ref{fig:diff}. The prior is set to $\pi_0=\pi_1=0.5$, which means that we do not favor any hypothesis before testing. From Fig~\ref{fig:diff}, as the range difference between two types increases, the $P_A$ and $P_D$ curves are associated with a more rapid change with respect to $N$. It suggests that if the qualities of the two types of sensors have a significant difference, it will be easier for the investigator to determine the accountability of the supplier within a fewer number of tests.

    \begin{figure}[ht]
            \centering
            \includegraphics[width=0.6\textwidth]{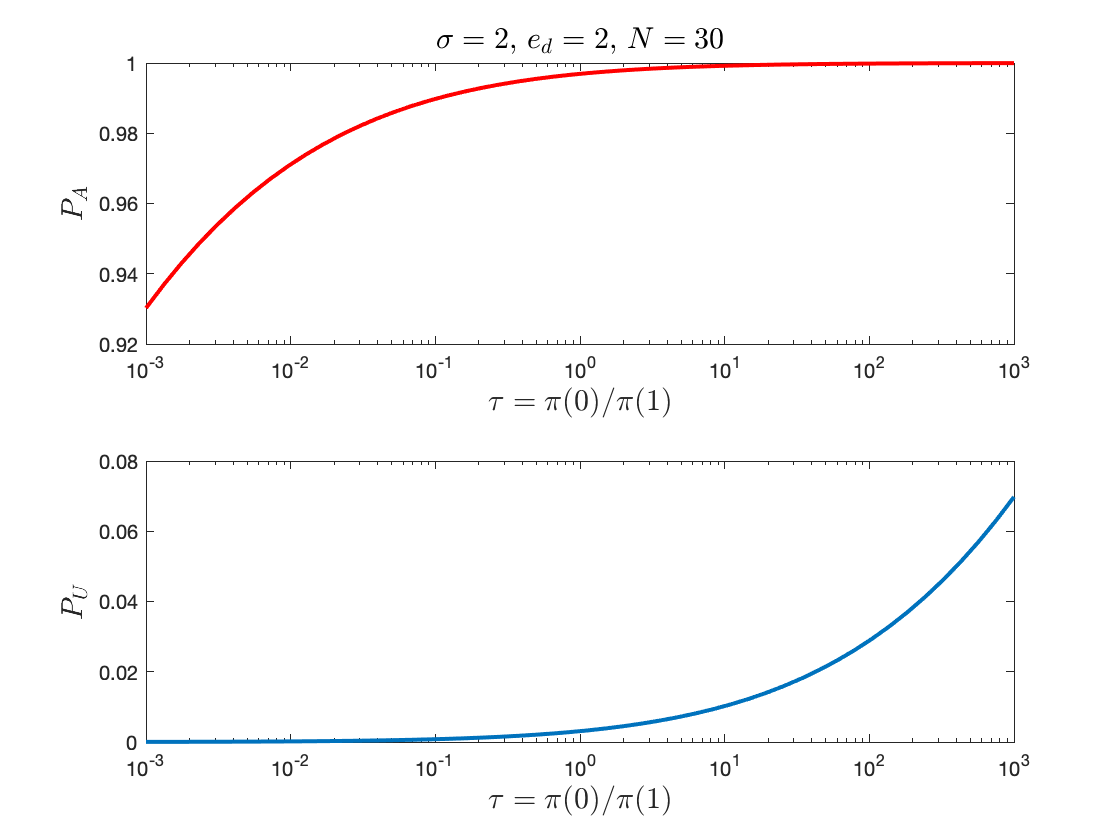}
            \caption{Impact of supplier's reputation ($\sigma=2$, $e_d=2$, $N=30$)}
            \label{fig:tau}
    \end{figure}

Fig.~\ref{fig:tau} displays the impact of supplier's reputation on the accountability estimation. The ratio $\tau = \pi_0/\pi_1$ represents the reputation of the supplier. A larger value of $\tau$ indicates that we have a strong belief the supplier is lying. Normally, we are more likely to suspect that the supplier with a bad reputation would be accountable for the incidents. As shown in Fig~\ref{fig:tau}, when we fix the testing environment, the accountability of supplier $P_A$ increases as $\tau$ increases. However, it should be noted that the wronged accountability $P_U$ increases as well. This is because the increase of $\tau$ will cause the testing threshold $\eta$ in LTR will increase, leading to a larger observation space $\mathcal{Y}_0$ where we classify the observations as $H_0$. Thus, both $P_A$ and $P_U$ will increase according to the definition. The wronged accountability misattributes the incident to the supplier when they should not be accountable. We will see more details about the trade-off between $P_A$ and $P_U$ in the following section.

\subsection{Investigation Performance}

\subsubsection{Accountability Receiver Operating Characteristic}
In the context of this ACC case study, we are interested in the relationship between accountability $P_A$ and wronged accountability $P_U$. as
\begin{align}
    P_A = \int_{\mathcal{Y}_0}f_m(\mathbf{y}|H_0)d\mathbf{y} = 1 - P_F \\
    P_U = \int_{\mathcal{Y}_0}f_m(\mathbf{y}|H_1)d\mathbf{y}= 1- P_D
\end{align}
Because of the symmetric property of the Gaussian $Q$ function, the ROC curve is invariant under this transformation. From equations (\ref{eq:pa}) and (\ref{eq:pu}), if we eliminate the parameter $\tau$, the relationship between $P_A$ and $P_U$ can be written as
\begin{align}
    P_U = Q(d - Q^{-1}(1-P_A))
\end{align}
The relationship between $P_A$ and $P_U$ is traced out as the threshold $\tau$ in LRT varies from $0$ to $\infty$. Note that this relationship depends on the variable $d=N^{1/2}e_d/\sigma$. We plot the ROC curve under different $d$ values in the following figure.

The slope of the AROC at point $\left(P_A(\tau),P_U(\tau)\right)$ is equal to the supplier's reputation $\tau$\cite{levy2008binary}. Ideally, we would like to conduct a hypothesis test such that $P_A$ is close to one and $P_U$ is close to zero. As we can see from the figure, the ROC curve approached the ideal test point when the value of $d$ increases. This result coincides with our aforementioned analyses. Increasing the number of test $N$, comparing sensor with larger sensor error $e_d$, and reducing the observation variance $\sigma$ can all increase the value of $d$, leading to a more reliable accountability test result.

    \begin{figure}[ht!]
            \centering
            \includegraphics[width=0.5\textwidth]{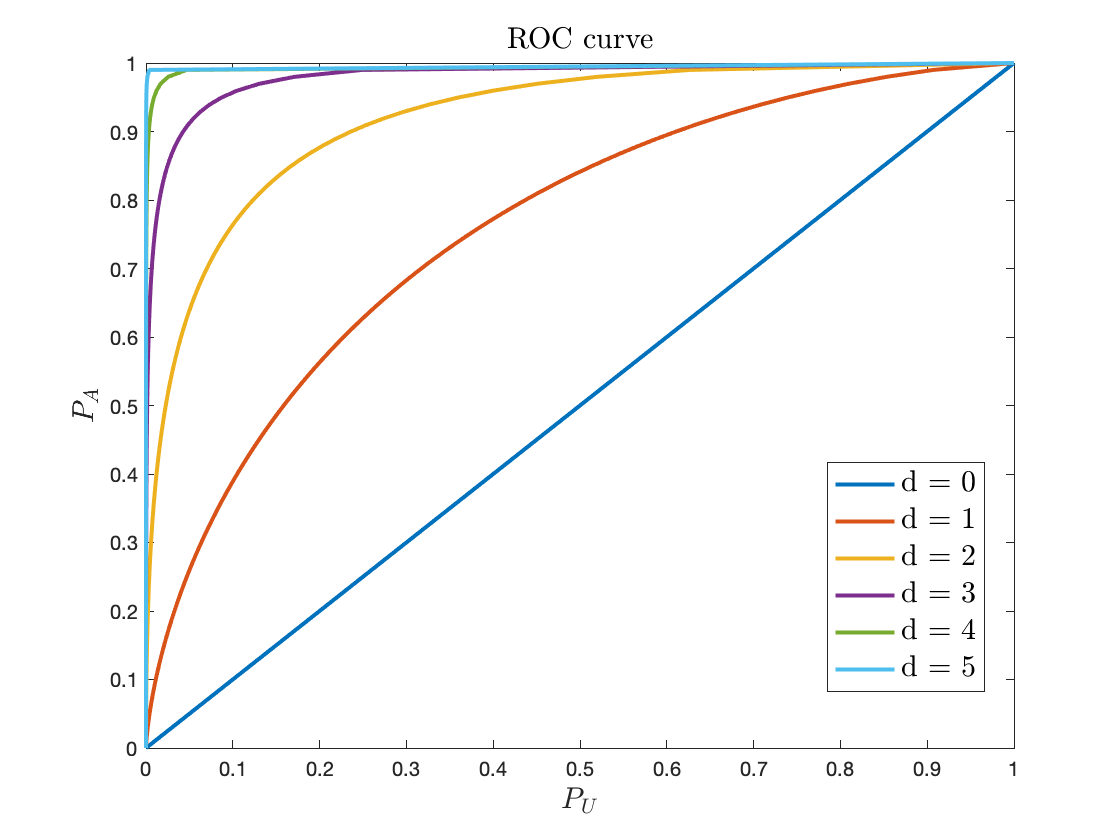}
            \caption{ROC curve under different $d$}
            \label{fig:roc}
    \end{figure}

\subsubsection{Area under the AROC curve}

\begin{figure}[ht!]
            \centering
            \includegraphics[width=0.5\textwidth]{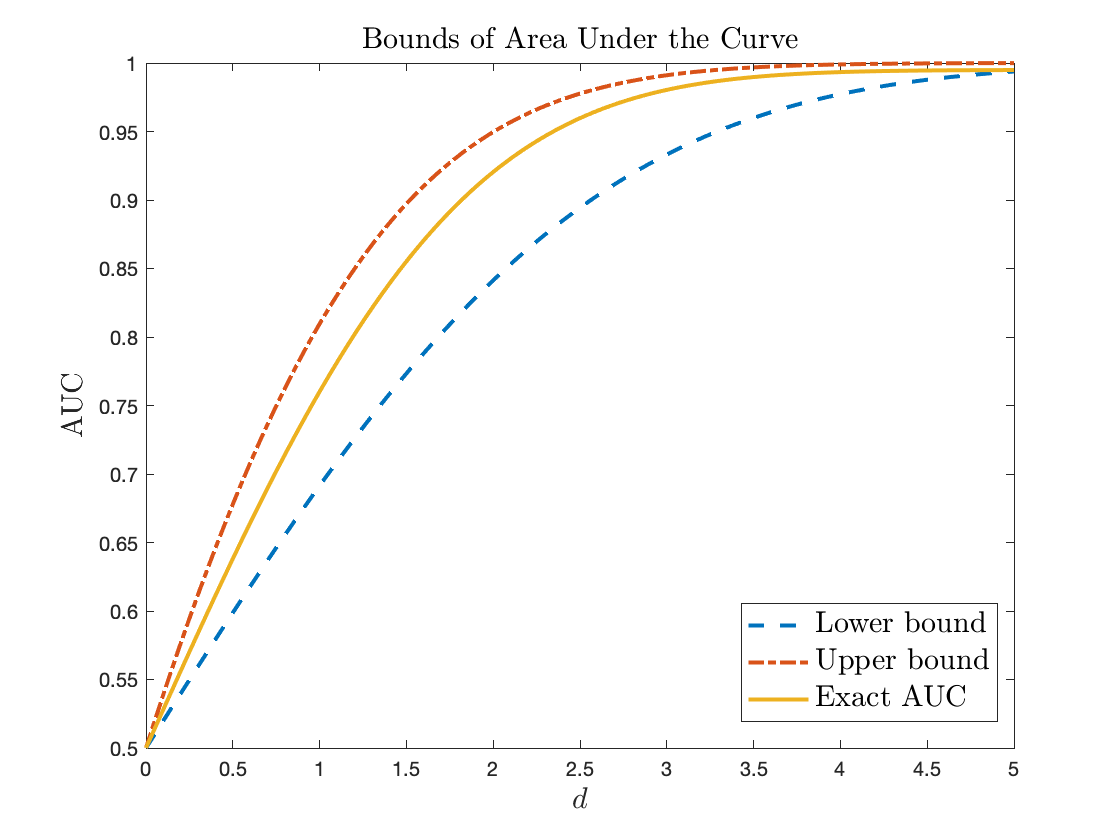}
            \caption{Bounds of AUC under different $d$}
            \label{fig:boundauc}
    \end{figure}

In the ACC sensor accountability testing case, the exact AUC value and its bounds with respect to $d$ are shown in Fig.~\ref{fig:boundauc}. From the figure, we can see that the performance of the hypothesis testing increases along with the value $d$. In fact, in testing with the Gaussian hypothesis, the value $d$ indicates the Chernoff distance between the two Gaussian distributions \cite{levy2008binary}. A larger value of $d$ means the distribution of $H_0$ and $H_1$ have less overlap, thus it is easier to separate between them. Since we have the exact expression of $P_e$, the bounds of AUC can be expressed as
\begin{align}
    1-Q\left(\frac{d}{2}\right) \leq AUC(d) \leq 1-2Q^2\left(\frac{d}{2}\right).
\end{align}

\newpage
\section{Case Study 2: Ransomware in IoT Supply Chain}

In this section, we provide a second case study of supplier accountability in smart home IoT under ransomware attacks. This example illustrates how we determine accountability in a supply chain and sophisticated systems involving different components.

\subsection{Background}

Ransomware is a type of malware that infects particular network entities to demand ransom. This kind of attack is becoming more prevalent nowadays with the fast development of IoT systems. The broad connections for IoT devices provide more security threats and vulnerabilities. Besides, the massive number of IoT devices increases the risk of getting infected by ransomware since any device could be the target. Indeed, the ransomware attack has caused significant economic losses in industrial domains. The estimated global damage from ransomware reaches \$20 billion in 2021 \cite{braue_2021}.

Smart home technologies integrate different IoT-enabled components to provide advanced services within the home environment. The components from different suppliers contribute to addressing various challenges to improve the quality of human life. However, their limited processing capabilities make them vulnerable to security threats \cite{geneiatakis2017security}, including ransomware. If the component in the home security system is taken controlled by the attacker, the end-user may face serious economic loss and privacy leakage. The user needs to determine which part of the IoT system should hold accountable for the accident. Our framework provides a way to mitigate the impact of ransomware by attributing the accident to a supplier of IoT devices.

\subsection{Smart Lock and Ransomware Attack}

Nowadays, smart home technologies have been widely accepted by individuals and organizations to improve home security. With the development of IoT and machine learning, the number of smart lock users are increasing in recent decades. Instead of physical keys, smart lock utilizes face recognition and/or fingerprint verification to achieve digital authentication. Most smart locks also are equipped with intruder alert and remote control when you are physically away from home. This innovation avoids the threats with cloneable physical keys and provides a front-line deterrent against potential intruders.  

While the smart lock offers convenience to homeowners, the transition towards digital control brings concerns over security in cyberspace. One potential threat is the ransomware attack. This type of attacks belongs to the family of Advanced Persistent Threats (APTs). A malicious attacker attack your smart home IoT system, lock the front door of your house, and request a ransom. The highly-connected feature of IoT provides the attacker multiple vulnerabilities as the entry point into the network. Once building a foothold in the network, the attacker moves laterally towards the target to achieve his goal, in this case, locking the door and denying legitimate access. Once compromised by ransomware, the dangling participle would be huge if someone under medical conditions is locked and requires immediate treatment. We may be discouraged by the fact that victims simply pay the ransom in many cases, and even the FBI once inadvertently mentioned paying the ransom if the network device is infected \cite{cartwright2019pay}.

    \begin{figure}[ht]
            \centering
            \includegraphics[width=0.5\textwidth]{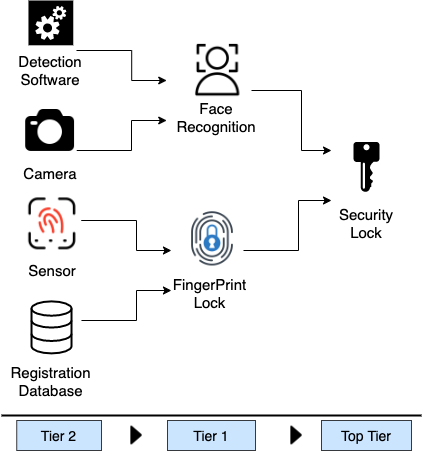}
            \caption{IoT supply chain related to security lock.}
            \label{fig:lock}
    \end{figure}

To mitigate the loss under such ransomware attacks, accountability investigation provides a way to check the responsibility of the IoT device supplier(s) regarding the attack. It is important for the investigator to find out the initial attack entry that poses a risk to the whole system. Due to the tiered structure of the supply chain, the accountability investigation needs to be constructed through a top-down layered tree analysis as shown in Fig.~\ref{fig:lock}. This structure helps the investigator to narrow down the search scope and determine the accountability of the suppliers among different supply chain tiers. More details will be provided in the following section.

\subsection{Accountability Investigation}
\subsubsection{Tier-1 Investigation}
Face recognition and fingerprint verification are two critical parts of smart lock authentication. The failure of the smart lock could be caused by the failure of one or both of the functions. In this case, the first step in accountability investigation is to determine whether the tier-1 suppliers of these two parts need to be accountable for the ransomware attack. As described in Sec.~\ref{sec:multiple}, this is corresponding to the accountability investigation of multiple suppliers.

Denote the supplier of face recognition technology as $i=1$ and the supplier of fingerprint verification technique as $i=2$. We assume that each supplier may have binary types $\theta_i \in\{0,1\}$.  $\theta_i=0$ means that the provided product operates normally and $\theta_i=1$ stands for malfunctioning. By default, each supplier sends a message $m_i = 0$ and guarantees the product functionality when signing the contract with the buyer. Thus, we can construct the following hypotheses as in Table~\ref{tab:4h}. Denote $h_i,i=\{1,2\}$ as the accountability of supplier $i$. $\hat{H}_0$ indicates that both parts are operating normally as reported; $\hat{H}_1/\hat{H}_2$ suggests that there be misinformation from one of the suppliers; $\hat{H}_3$ means both suppliers need to hold accountable for the ransomware attack.
\renewcommand{\arraystretch}{1.3}{
\begin{table}[ht]
\begin{center}
\begin{tabular}{ | c|c| c | } 
  \hline
  Hypothesis & $h_1 = \mathds{1}(\theta_1 \neq 0)$ & $h_2 = \mathds{1}(\theta_2 \neq 0)$ \\ 
  \hline
  $\hat{H}_0$ & 0 & 0 \\ 
  \hline
  $\hat{H}_1$ & 0 & 1 \\ 
  \hline
  $\hat{H}_2$ & 1 & 0 \\ 
  \hline
  $\hat{H}_3$ & 1 & 1 \\ 
  \hline
\end{tabular}
\caption{Four hypotheses in accountability investigation.}
\label{tab:4h}
\end{center}
\end{table}}

\begin{figure}[t]
            \centering
            \includegraphics[width=0.7\textwidth]{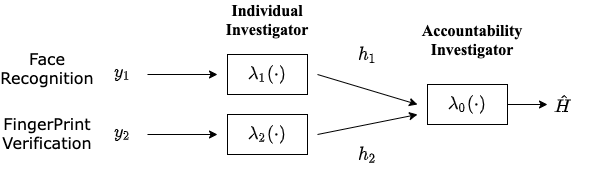}
            \caption{Decentralized tier-1 accountability investigation. }
            \label{fig:tier1}
    \end{figure}

Instead of looking into the joint performance of the two components, it is practical to conduct independent decentralized investigations into each of the suppliers as shown in Fig~\ref{fig:tier1}. We take the face recognition system $h_1$ for example. The investigation of the fingerprint verification $h_2$ can be conducted in the same manner.
Suppose the normal operating face recognition system can correctly detect the registered identity with $\mu_0 = 9\%$ accuracy. If this system is destructed by the ransomware attacker, we would expect a lower identification accuracy, i.e. $\mu_1 <\mu_0$. To investigate the accountability of the face recognition system, we design the following testing scenarios. 
On each trial, different photos of registered faces are displayed randomly in front of the device. The performance $y_i\in \{0,1\}$ at each trial is an indicator of the testing results, where $y_i=1$ represents correct identification and $y_i=0$ otherwise. Let $Y^N = \{y_1,y_2,\dots,y_N\}$ be a sequence of independent and identically distributed trials, we consider the following hypotheses for accountability testing. For each trail $1\leq i\leq N$, 
        \begin{align*}
            H_0: \, y_i \sim \text{Bern}\left(0.9\right),\qquad H_1: \, y_i \sim \text{Bern}\left(\mu_1\right),
        \end{align*}
where $\mu_1<\mu_0=0.95$. Bernoulli distribution is a natural model to describe events with Boolean-valued outcomes under certain success probability. In this hypothesis model, $H_0$ indicates that the face recognition system operates normally with $90\%$ detection accuracy on average. $H_1$ suggests a degraded identification accuracy. This investigation aims to find out whether hypothesis $H_1$ should be established based on the system performance.

One limitation of Bayesian tests as described in Sec.\ref{sec:aeb} is their reliance on the prior knowledge $\pi$, i.e., the reputation of the supplier, and costs assigned to different decision errors. The choice of decision cost depends on the nature of the problem, but the prior probabilities must be known. In many applications, the prior knowledge may not be obtained precisely; thus, the correct value of the threshold in LRT is unknown. In the ransomware case study, the misinformation between the supplier and buyer may be unintended. It is challenging to determine the probability $\pi_1$ that the supplier is compromised by the attacker. It is natural to consider alternative tests that can achieve desired detection results without such prior knowledge.

Neyman and Pearson \cite{neyman1933ix} formulated a test $\lambda$ that maximizes the correct detection probability $P_A(\lambda)$ (accountability) while ensuring the false-alarm probability $P_U(\lambda)$ (wronged accountability) is subject to a upper bound constraint $\alpha$. This can be formulated as
\begin{equation}
\begin{aligned}
\max_{\lambda} \quad &  P_A(\lambda)= \int_{\mathcal{Y}_1} f_m(Y^N|H_0)dy^N,\\
\textrm{s.t.} \quad & P_U(\lambda)= \int_{\mathcal{Y}_1} f_m(Y^N|H_1)dy^N \leq \alpha.\\
\end{aligned}
\end{equation}
This constrained optimization problem requires no prior knowledge about reputation and decision cost function. The only parameter that needs specification is the maximum acceptable wronged accountability $\alpha$. A classic result due to Neyman and Pearson shows that the optimal solution to this type of investigation is a likelihood ratio test (LRT).

\begin{lemma}[Neyman-Pearson]
Consider the likelihood ratio test in (\ref{LRT}) with $\tau_k>0$ chosen so that $P_U(\tau_k)=\alpha$. There does not exist another test $\lambda$ such that $P_U(\lambda)\leq \alpha$ and $P_A(\lambda)\geq P_A(\tau_k)$. Hence, the LRT is the most powerful test with false-alarm probability $P_U(\lambda)$ less than or equal to $\alpha$.
\end{lemma}

In the accountability investigation of the face recognition system, both hypotheses admit a Bernoulli distribution. The likelihood ratio is given by 
\begin{align*}
    L(Y^k) = \frac{\prod_{i=1}^N \mu_1^{y_i}(1-\mu_1)^{1-y_i}}{\prod_{i=1}^N \mu_0^{y_i}(1-\mu_0)^{1-y_i}} = \left(\frac{1-\mu_0}{1-\mu_1}\right)^N\left(\frac{\mu_0(1-\mu_1)}{\mu_1(1-\mu_0)}\right)^{\sum_{i=1}^N y_i}.
\end{align*}

The sufficient statistics of such testing will be the sum of all performance results $S = \sum_{i=1}^N y_i$. According to Neyman-Pearson lemma, the most powerful test will hold the supplier accountable if $S<\lambda$ for a constant threshold $\lambda$.
\begin{align*}
    S = \sum_{i=1}^N y_i\mathop{\gtreqless}_{H_1}^{H_0} \lambda
\end{align*}
Under $H_0$, the detection accuracy is on average $\mu_0$, and $S$ admits to a binomial distribution, $S\sim$ Binomial$(N,\mu_0)$.
To ensure $P_U(\lambda)=\alpha$, the threshold $\lambda$ is chosen to be the $\alpha$ quantile of the
Binomial$(N,\mu_0)$ distribution. 
\begin{align*}
    \lambda = Q(\alpha)\,=\,\inf\left\{ x\in \mathbb{R} : \alpha \le F_S(x) \right\},
\end{align*}
where $F_S(x)$ is the cumulative distribution function of random variable $S$. Note that as this is a discrete distribution, it may not be possible to get the exact $\alpha$ and $\lambda$ desired. One way to address this problem is to increase the total number of trials $N$ and approximate the binomial with a Gaussian distribution according to the central limit theorem.
 
    \begin{figure}[ht]
            \centering
            \includegraphics[width=0.6\textwidth]{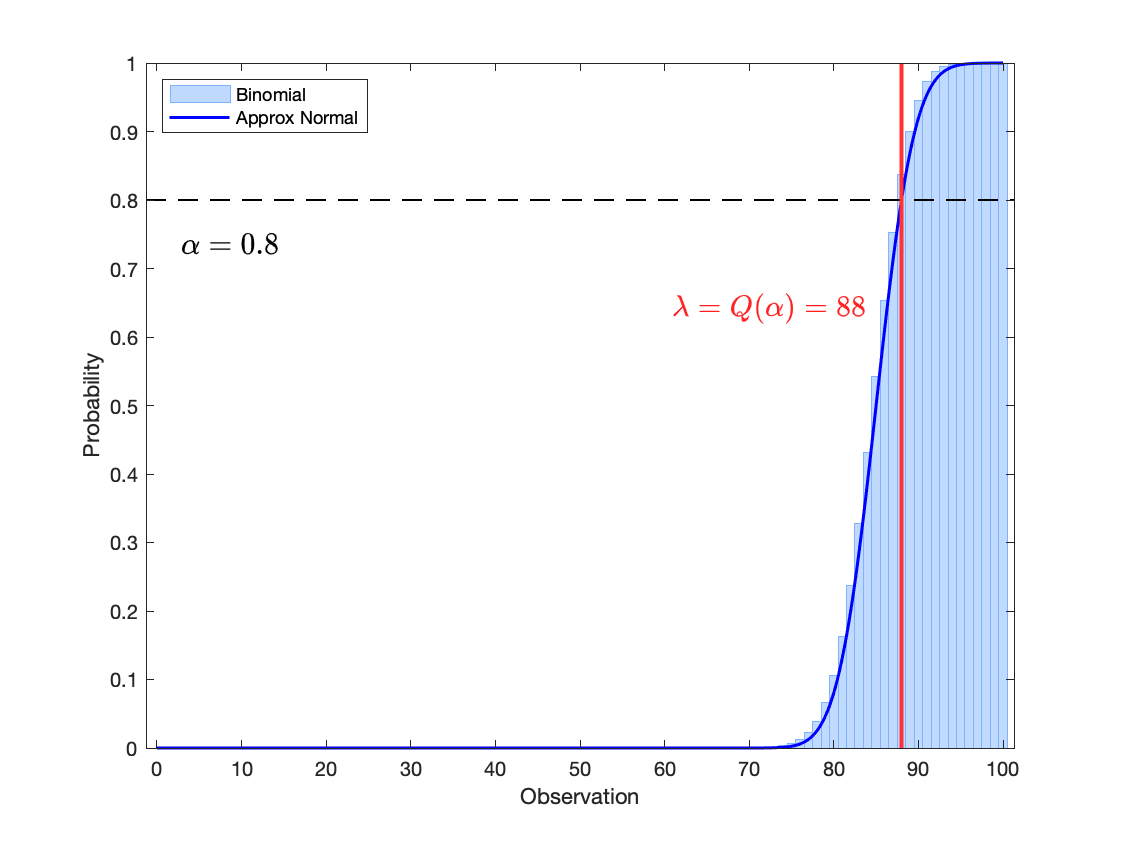}
            \caption{Neyman-Pearson test result for tier-1 investigation. }
            \label{fig:alpha}
    \end{figure}

In the IoT ransomware attack case, the changes made by the stealthy attacker often remains unknown even after investigations. Thus, it is hard to determine identification accuracy $\mu_1$ after the attack and find the exact performance distribution under hypothesis $H_1$. We can only assume that the attack results in a degraded identification accuracy as $\mu_1<\mu_0$. Neyman-Pearson test provides a way to investigate the accountability of the supplier with limited prior knowledge. It guarantees that the correct detection probability $P_A$ is  maximized under the false-alarm constraint $P_U\leq \alpha$. In the context of the IoT supply chain attack, Neyman-Pearson test paves the way for the buyer to investigate the accountability of the supplier with limited information.

\subsubsection{Multi-stage Accountability Investigation}

The tier-1 investigation examines the accountability of each tier-1 supplier. However, due to the layered structure of the IoT supply chain and the sophisticated feature of the ransomware attack, the true cause of the attack may lie in the suppliers in the subordinate tiers. Tier-1 suppliers can further attribute the malfunction to their suppliers following a similar fashion. A top-down layered investigation is needed if we would find out the origin of the attack and obtain a holistic view of the entire supply chain. This is called a multi-stage accountability investigation.

For instance, if the face recognition system should hold accountable for the attack according to the tier-1 investigation, the supplier could further investigate the components that the system consists of. There may exist different types of vulnerabilities in the components that are provided by tier-2 suppliers. The attacker could break into the system by compromising the ill-protected camera and further penetrating into the system. Another possibility is that adversaries against face recognition are performed at the detection software. If the latter case holds true, the detection software provider can further check which part of the software is malfunctioning. Face recognition attacks can be performed at the database, the predefined algorithm parameters, the communication channels, etc. The multi-stage accountability investigation aims to further figure out which among the vulnerabilities is the underlying cause of the attack.

    \begin{figure}[ht]
            \centering
            \includegraphics[width=0.85\textwidth]{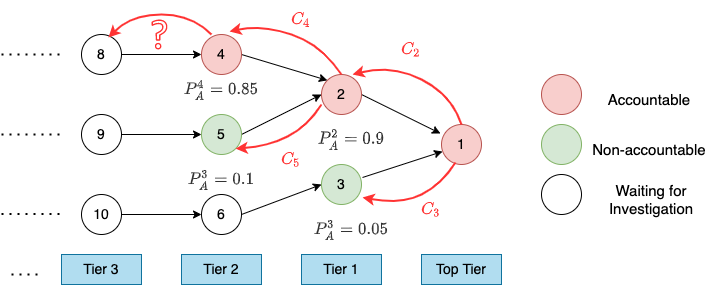}
            \caption{Multi-stage accountability investigation. }
            \label{fig:mult}
    \end{figure}

To analyze the accountability of the involved suppliers at each tier, we view the supply chain as a directed graph as shown in Fig.~\ref{fig:mult}. The arrows in the graph indicate the procurement relationship. Multi-stage accountability starts from the top tier node, the final product. The accountability investigation on each supplier $i$ produces  accountability $P_A^i$ subject to an investigation cost $C_i$. Whether a supplier is accountable depends on the comparison between $P_A^i$ and selected threshold $\epsilon\in(0,1)$. We call a supplier accountable if $P_A^i>\epsilon$.

If the current supplier is determined to be non-accountable ($P_A^i<\epsilon$), there is no need to continue investigation among its suppliers. In the ransomware example, if we determine that the face recognition system solely should hold accountable in the tier-1 investigation, there is no need to conduct an accountability check for the suppliers related to the fingerprint verification system. Deductive reasoning helps reduce the investigation efforts on unrelated system components and focus on the ones that attribute the accident. It provides a way to prioritize the contributors leading to the top event. 

It should be noted that the product design of each sub-system can also be the cause of the vulnerability that exposes the system to threats. This brings up the question that how deep we should investigate during the process. Suppose the total investigation budget is $B$. The investigator needs to decide whether to continue the investigation or simply stop and replace the component. Replacement will be a better choice if the remaining budget cannot support further investigation as
\begin{align*}
    B - \sum_{i\in\mathcal{I}}C_i \leq C_{\text{next}},
\end{align*}
where $\mathcal{I}$ is the set of investigated suppliers and $ C_{\text{next}}$ is the investigation cost of the next supplier. The trade-off between investigation and replacement may be another dimension to be considered when conducting multi-stage accountability investigations.

Multi-stage accountability investigation is an iterative analysis process to find the cause of the accident. The layered approach provides a way to understand how the system fails, identify the vulnerabilities in the IoT supply chain, and determine the accountability of any supplier. It also creates the foundation for any further analysis and evaluation. If the structure of the supply chain has been upgraded (e.g., component replacement), it can provide a set of steps to design quality tests and maintenance procedures.

\section{Compliance and Cyber Insurance} \label{sec:com+insurance}

\subsection{Compliance Modeling}

The description $m\in M$ from the supplier to the buyer is a self-reporting mechanism that requires the vendors to disclose information about their products so that the buyers can use the NIST standards to check their compliance before they are integrated into IoT systems. The procured products have to comply with the business or mission, organization-specific requirements, the operational environment, risk appetite, and risk tolerance \cite{Boyens_2015}. Security requirements are an important component of compliance. They are imposed by not only the developers in the private sectors to provide information and quality assurance but also the law, which aims to protect the nation from cyber-attacks. 

Recent legislation has been signed into law requiring IoT devices purchased with government money to comply with security standards \cite{wednesday_kovaleski_2020}. The Internet of Things Cybersecurity Act of 2020 \cite{kelly_2020} requires NIST to ``develop and publish under section 20 of NIST Act (15 U.S.C. 278g-3) standards and guidelines for the federal government on the appropriate use and management of Internet of Things devices owned or controlled by  an agency and connected to information systems owned or controlled by an agency, including minimum information security requirements for managing cybersecurity risks associated with such device." All IoT devices connected to IT systems owned or controlled by a federal agency must conform to NIST standards by September 4, 2021. 

The Biden executive order of May 12, 2021 \cite{wh_2021} demands that ``the federal government must bring to bear the full scope of its authorities and resources to protect and secure its computer systems, whether they are cloud-based, on-premises, or hybrid." The scope of protection and security must include systems that process data (information technology (IT)) and those that run the vital machinery that ensures our safety (operational technology (OT))." The executive order requires full NIST compliance. The focus of the new rules is on IoT systems that support information technologies, e.g., the power and cooling systems, such as uninterruptible power supplies (UPSs), power distribution units (PDUs), and computer room air conditioners and air handlers (CRAC \& CRAH) that support networks, servers, and data centers on the property of federal agencies, building management systems (BMS), and data center infrastructure management systems (DCIM).

Besides the federal regulations, supply contracts are also useful to secure systems installed by suppliers. The suppliers need to be informed of your security requirements and standards. You can check whether the proposed or delivered products or services comply with them. The contracts also play an important role in accountability. The penalty can be enforced by contracts once non-compliance of the services is found by the buyer, which has been discussed in the earlier section.

We can use formal methods to check whether the attributes in $m$ satisfy the requirements that are coded into logical formulae $f$. The product is compliant if $m \models p$, the description satisfies the specifications; otherwise, it is not. There are well-established tools that can be used to efficiently solve this satisfiability problem. For example, the compliance problem can be formulated as a satisfiability modulo theories (SMT) problem, which can be solved using a formalized approach and many solvers. PRISM is another tool that enables probabilistic modeling and checking of systems. Under the assumption that the reporting of $m$ truthfully describes the product, i.e., $m=\theta$, a compliant buyer or system will not acquire from suppliers that do not satisfy the requirement. In other words, $a=0$ if $m \not\models p$.

\subsection{Contract Design}

There are two economic-level solutions. One is the mechanism design between the buyer and the supplier to induce $m=\theta$. To achieve this, we would need to create incentives for the supplier to truthfully reveal $\theta$. This would rely on the design of a certain form of penalty as a credible threat. One of such penalties is through the contract. The contract between the supplier and the buyer would include a penalty once the supplier is accountable. The contract will be effective only when the buyer decides to purchase the product $a=1$, which happens with probability $\alpha(m) = Pr(a=1|m)$. We consider the following utility function of the supplier, $U_S:\Theta\times\mathcal{M}\mapsto\mathbb{R}$, given by 
\begin{align}
U_S(\theta, m):= & \mathbb{E}_\alpha\left[J_S(\theta, m) - \mathbb{E}_{P^m_A}\left[C_S(\theta, m)\right]\right].
\end{align}
Here, $J_S:\Theta\times\mathcal{M}\mapsto\mathbb{R}$ is the profit of the supplier if he reports $m\in\mathcal{M}$ when the true type is $\theta\in\Theta$ and under the procurement decision. The second term in the utility function is the average penalty $C_S:\Theta\times\mathcal{M}\mapsto\mathbb{R}$ for the supplier if he is held accountable. The probability of being accountable is given by $P^m_A$ in Def.~\ref{accountability} based on the received message $m$. It is clear that the penalty depends on $\theta$ and $m$.

We call a supplier is incentive-compatible if 
\begin{align}
U_S(\theta, \theta) \geq U_S(\theta, m), \ \ \textrm{for~all} \ m\in M \tag{$IC_S$}.
\end{align} 

An incentive-compatible supplier does not have incentives to misreport what he knows when he is held accountable for his actions.  Note that to achieve this, we assume that the purchase rule and accountability testing scheme are revealed to the supplier through the contract. The ($IC_S$) condition gives a natural constraint when designing a procurement contract. However, the challenge is that the profit function $J_S$ and the type space of the suppliers are often unknown to the acquirer and they need to be conjectured or learned from experience or data.

We call a supplier is individually rational if 
\begin{align}
    U_S(\theta,m)\geq 0, \ \ \textrm{for~all} \ m\in M, m\neq \theta \tag{$IR_S$}
\end{align}

The ($IR_S$) constraint ensures the supplier will benefit from participating in the contract. This requires the buyer to design the penalty carefully so that the expected profit of the supplier is non-negative.

\subsubsection*{Example: Autonomous truck platooning}

If we take a closer look at the utility function of the supplier, it can be further expressed as
\begin{align}
    U_S(\theta, m)= \alpha(m)\cdot \left[J_S(\theta, m) - C_S(\theta, m)\cdot P^m_A\right].
\label{eq:ic2}
\end{align}
The goal of contract design is to assign an appropriate penalty $C_S$ for the supplier if they need to be held accountable for the accident. The first consideration comes from the ($IR_s$) constraints. This set of constraints suggests that we should not assign a penalty that exceeds the expected profit.

The ($IC_S$) constraints are automatically satisfied when the supplier truthfully report $m=\theta$. Consider the autonomous truck platooning example as described in Sec.~\ref{sec:aeb} with the binary sensor type space, i.e., $\Theta=M=\{0,1\}$. The contract designer need to meet the following constraints
        \begin{align}
            \alpha(1)\left(J_S^{11}-P_A^1 C_S^{11}\right) \geq \alpha(0)\left(J_S^{10}-P_A^0C_S^{10}\right) \label{ic1}\\
            \alpha(0)\left(J_S^{00}-P_A^0 C_S^{00}\right) \geq \alpha(1)\left(J_S^{01}-P_A^1C_S^{01}\right)  \label{ic2}
        \end{align}
where we denote the profit of supplier with true type $\theta$ who sends message $m$ as $J_S^{\theta,m}$, and the penalty for such supplier as $C_S^{\theta,m}$.

From the contract designer's viewpoint, the profit of the supplier $J_S^{\theta,m}$ is beyond his control. This value is determined by the production cost and economical nature of the system. In the ACC system, $\theta=1$ is the product type corresponding to the system design. It is natural to assume that the sensor supplier with true type $\theta=1$ will make more profit when he truthfully reports, as $J_S^{11}>J_S^{10}$. Similarly, we can assume misinformation will bring more profit for the supplier with $\theta=0$, as $J_S^{00}<J_S^{01}$.

In terms of misinformation penalty, it is incentive to penalize more on the supplier who fails to truthfully  report, as $C_S^{\theta,\theta}<C_S^{\theta, m}$, for every $m\neq \theta$. If we expect same purchasing policy $\alpha(m)$ and accountability $P_A^m = P_A$ are the same for both messages $m\in\{0,1\}$, constraint (\ref{ic1}) will be automatically satisfied and constraint (\ref{ic2})  will be reduced to 
\begin{align}
            J_S^{01}-J_S^{00}\leq P_A (C_S^{01}-C_S^{00}).
\end{align}
This indicates for the supplier $\theta=0$ who has the incentive to misinform the buyer, the expected extra penalties brings to the supplier through contract need to exceed the extra profit generated from the untruthful report. The result coincides with the intuition that the contract needs to be designed with incentive capability.

\begin{figure}[ht]
\centering     
\subfigure[Feasible Region for $\theta=0$]{\label{fig:feac0}\includegraphics[width=0.48\textwidth]{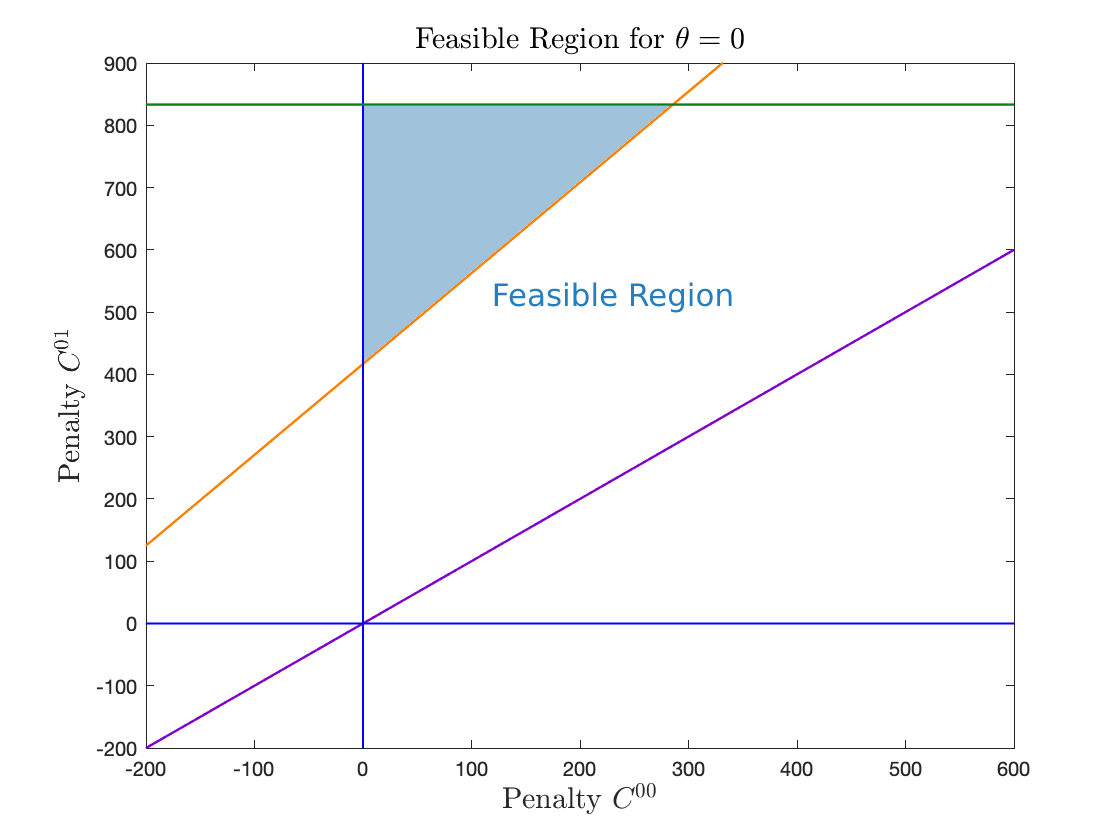}}
\subfigure[Feasible Region for $\theta=1$]{\label{fig:feac1}\includegraphics[width=0.48\textwidth]{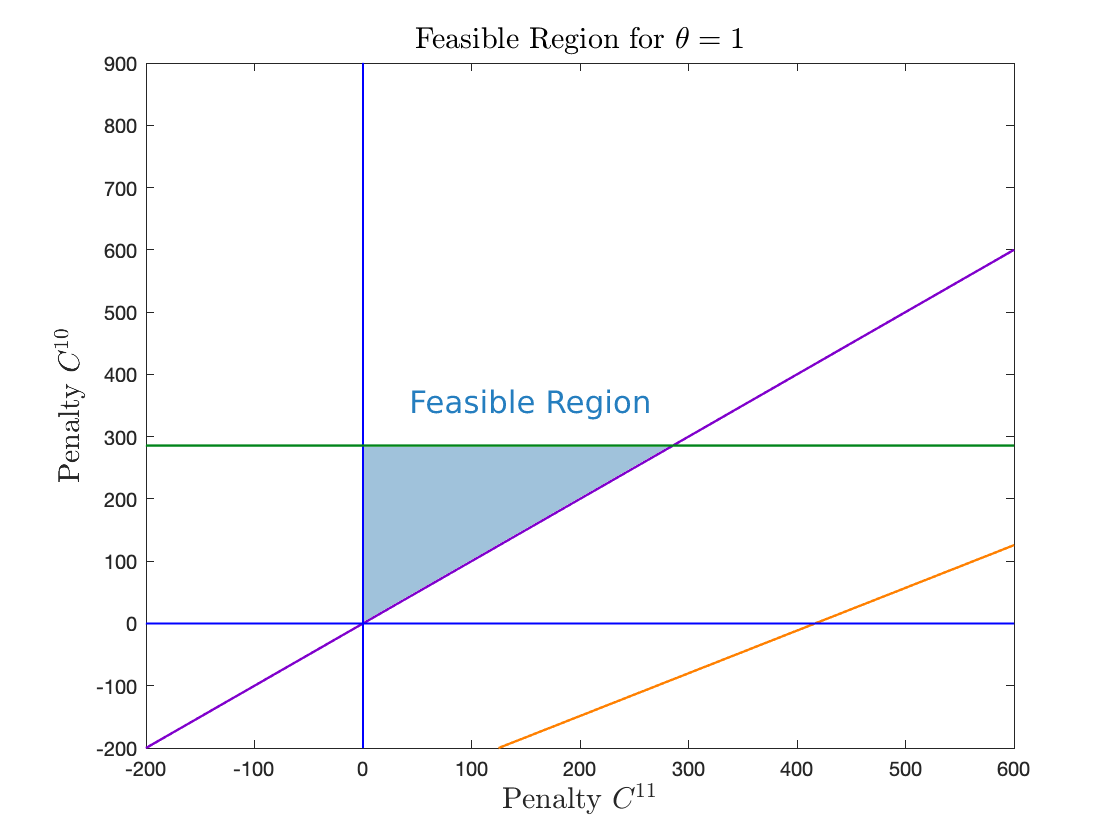}}
\caption{Feasible penalties under constraints.}
\label{fig:feac}
\end{figure}

For automakers looking at production, the prices of lidar sensors need to be cost-effective for automotive ACC use. Ranging sensors with greater abilities will be sold for higher prices. It is reported that Lidar suppliers manage to reduce the single-unit samples price to \$250 in large volumes \cite{hecht2018lidar}. In the ACC supplier example, consider the following values:
\begin{align*}
    J_S^{11} = J_S^{01} = 250; \quad &J_S^{00} = J_S^{10} = 200;\quad
    \alpha(1) = 0.8, \alpha(0)=0.5; \quad P_A^1 = 0.3, P_A^0=0.7.
\end{align*}
We arrive at the following constraints for contract penalty design for the supplier:

\begin{align}
    &0.8*(250-0.3*C_S^{11}) \geq 0.5*(200 - 0.7*C_S^{10}), \notag\\
    &0.5*(200-0.7*C_S^{00}) \geq 0.8*(250 - 0.3*C_S^{01}), \tag{$IC_S$}\\
    &0.5*(200 - 0.7*C_S^{10})\geq 0,\notag\\
    &0.8*(250 - 0.3*C_S^{01})\geq 0,\tag{$IR_S$}\\
    &C_S^{00}<C_S^{01},\quad C_S^{11}<C_S^{10}. \notag
\end{align}
By solving the feasible region of penalty under constraints as in Fig.~\ref{fig:feac}, the contract designer can select the proper penalties for the supplier and help avoid misinformation.

\subsection{Cyber Insurance}

\subsubsection{Background Introduction}

In spite of the wide applications of cyber-physical systems, the cyber risks within the IoT supply chain are considered to be the most challenging problem to handle. Cyber insurance is the last resort for resilience to mitigate the loss of performance. It is an important risk management tool that transfers the risks of the buyer to a third party, i.e., an insurer. Victims of a cyber attack can reduce their financial losses and quickly recover to restore their business operations.  According to the cyber insurance report released by the National Association of Insurance Commissioners (NAIC) \cite{naci2020}, the cybersecurity insurance market in 2020 is roughly \$4.1 billion reflecting an increase of 29.1\% from the prior year. This scheme particularly benefits small and medium-size businesses that cannot afford a major investment in cyber protection.

Unlike traditional insurance policies, cyber insurance reimburses the buyer for the loss incurred by data breaches, malware infections, or other cyberattacks in which the insured entity was at fault. An incentive-compatible cyber insurance policy could help reduce the number of successful cyber attacks by incentivizing the adoption of preventative measures in return for more coverage \cite{cashell2004economic,majuca2006evolution}. It can be served as an indicator of the quality of security protection. Besides, it is believed that cyber insurance can induce greater social welfare and encourage more comprehensive policies regarding cyber security\cite{marotta2017cyber}.

Various frameworks have been proposed to study cyber insurance from different perspectives. Pal et al., studies the economic impact of cyber insurance by proposing a supply-demand model. Their work showed that cyber insurance with client contract discrimination can improve network security \cite{pal2014will}. Böhme et al. proposed several market models to understand the information
asymmetries between defenders and insurers \cite{bohme2010modeling}. Radanliev et al. built a new impact assessment model of IoT cyber risk to better estimate cyber insurance \cite{radanliev2018analysing}. In our framework, we will focus on the cyber insurance policy within the IoT supply chain and understand the impact of accountability investigation on cyber insurance.

\subsubsection{Insurance Policy Design}

Typically, the cyber insurance contract consists of the premium price and the coverage rate. The key challenge in insurance policy design lies in the difficulty of risk evaluation due to the complex structure of the cyber-physical systems. An insurer can make two separate contracts with the supplier or/and the buyer. The loss of the buyer would be compensated by the insurer when an accident or a disruption occurs. The loss of the supplier due to accountability could be insured as well. In this section, we focus on the insurance contract between an insurer and a buyer.

    \begin{figure}[ht]
            \centering
            \includegraphics[width=0.85\textwidth]{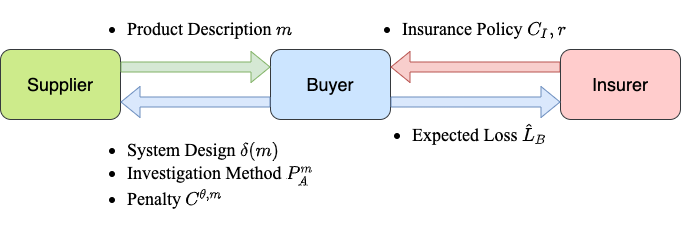}
            \caption{Information exchange between the insurer, buyer and supplier. }
    \end{figure}

The contract is composed of the premium and the coverage of the losses. Let $C_I\in \mathbb{R}$ be the premium charged by an insurer and the coverage is modeled by the percentage $r\in(0,1]$. They are decision variables that are determined by the insurer. A buyer has incentives to participate in the insurance if the average utility under the coverage is higher than the one without coverage. To quantitatively capture it, we specify the loss or payoff function of the buyer $J_B$, given by

\begin{equation}\label{eq:jb}
J_B(m , \delta): = (1-r) \hat{L}_B(m, \delta(m); \theta) + C_B(m) + C_I.
\end{equation}
Here, the first term $\hat{L}_B$ is the average loss of performance, which is the difference between the true and the anticipated performances. The cyber insurance will cover the $r$ portion of the risk. Hence the residual loss is $(1-r)$ of the losses. The insurance can completely compensate for the loss of the performance when $r=1$. The second term is $C_B(m)$ is the cost of procurement of the product and $C_I$ is the premium paid by the buyer.

In this framework, we focus on the potential loss due to the misinformation from the supplier who cannot be held accountable due to the limitation of accountability investigation. According to the investigation, if the supplier should be held accountable for the malfunctioning of the system, the loss of performance should be compensated by the supplier. However, if the investigation cannot hold the supplier accountable, the risk will be transferred to the third party under the insurance contract. The latter case occurs with probability $1-P_A^m$, \textit{the probability of unaccountable}. Thus, the loss of performance can be viewed as a stochastic variable $l_B$

\begin{equation}
    l_B(m, \delta(m); \theta)=
    \begin{cases}
        U_B(m, \delta(m)) - U_B(\theta, \delta(m)) & \text{w.p.  } 1-P^m_A, \\
        0 & \text{w.p.  } P^m_A,
    \end{cases}
    \label{eq:loss_b}
\end{equation}
where $U_B(\theta,\delta(m))$ is the performance utility measure under the design $\delta(m)$ and the true product quality $\theta$. We assume that the true performance $U_B(\theta, \delta(m))$ is at best the same as the anticipated performance when $m=\theta$, i.e. $U_B(m, \delta(m))$. When misinformation occurs, there will be a positive loss of performance; when the supplier truthfully report, the true performance coincide with anticipated one and the loss is zero; in other words, the expected loss of performance
\begin{align}
    \hat{L}_B = (1-P_A^m)\Delta U_B \geq 0,
\end{align}
where we denote the difference in performance measure as $\Delta U_B$.

One critical aspect of cyber insurance is the bias from insurance buyers. Humans will hold biased recognition concerning losses and risks, which can lead to different decisions compared to completely rational ones. Agents are often risk-averse, which means they prefer lower returns with known risks rather than higher returns with unknown risks. In terms of the expected losses $\hat{L}_B$, economic literature commonly imposes the following functions for a risk-averse agent.

\begin{itemize}
    \item Constant Absolute Risk Aversion (CARA) \cite{bohme2010modeling}:
    \begin{align}\label{eq:cara}
        \phi(x) = \frac{e^{\beta x}}{\beta},
    \end{align}
    where the parameter $\beta\leq 1$ is the absolute risk aversion coefficient, measuring the degree of risk aversion that is implicit in the utility function. The biased expected loss in this case is 
    \begin{align}
        \Phi(\hat{L}_B) = (1 - P_A^m) \phi(\Delta U_B),
    \end{align}
    
    \item Prospect Theory (PT) \cite{kahneman2013prospect}:
    \begin{equation}
    \label{eq:prospect_theory}
    \phi(x) = \begin{cases}
        x^\beta & x \geq 0 \\ 
        -\lambda (-x)^\beta & x < 0
    \end{cases},
    \quad 
    w(p) = \frac{p^\zeta}{p^\zeta + (1-p)^\zeta},
    \end{equation}
    where $\phi(x)$ and $w(p)$ are biased utility and weighted probability, respectively, and $\lambda, \beta, \zeta$ are prospect parameters with loss aversion implying $\lambda>1$.  In general, PT shows that people are more averse to losses and less sensitive to gains; people inflate the belief for rare events and deflate for high-probability ones. 
    The biased expected loss in this case is 
    \begin{align}
        \Phi(\hat{L}_B) = w(1 - P_A^m) \phi(\Delta U_B),
    \end{align}
\end{itemize}

For these types of buyer, we should replace the average loss $\hat{L}_B$ in equation (\ref{eq:jb}) with the biased expectation $\Phi(\hat{L}_B)$. The risk-averse buyer has an incentive to purchase cyber insurance if the expected cost under insurance is lower than the one without insurance: 
\begin{align}
(1-r) \Phi(\hat{L}_B) + C_B(m) + C_I \leq \Phi(\hat{L}_B) + C_B(m).  \tag{$IR_B$}
\end{align}
Note that we assume that the utility of the buyer does not include the penalty payment from the procurement contract and assume that the procurement does not involve an accountability contract. If so, we need to design the procurement contract and the insurance contract jointly as they are interdependent. 

The mechanism design problem of the insurer is to determine the optimal premium rate $C_I$ and the coverage $r$ to maximize his profit. The insurer provides insurance only when the the profit is non-negative. Thus, we have the following constraint.
\begin{align}
J_I := C_I - r\cdot \hat{L}_B \geq 0  \tag{$IR_I$}
\end{align}
We assume that the insurer is rational and risk-neutral so that they use the accurate value of the expected loss of the system when making decisions. The insurer solves the following optimization problem:
\begin{equation}
\begin{aligned}
\max_{r,\, C_I} \quad & J_I = C_I - r\cdot \hat{L}_B &\\
\textrm{s.t.} \quad & (1-r) \Phi(\hat{L}_B)+ C_I \leq \Phi(\hat{L}_B) &   \quad\text{($IR_B$)}\\
\quad & C_I - r\cdot \hat{L}_B \geq 0  & \quad \text{($IR_I$)}\\
& r\in (0,1]\\
& C_I\in \mathbb{R}^+
\end{aligned}
\end{equation}

Combining the individual rationality constraints ($IR_B$) and ($IR_I$) with the biased utility function, we arrive at the following proposition.
\begin{proposition} 
The insurance contract is established between the insurer and the buyer if the premium $C_I\in\mathbb{R}^+$ and the coverage level $r\in(0,1]$ satisfy
\begin{align}
    \hat{L}_B\leq \frac{C_I}{r}\leq \Phi(\hat{L}_B )
\end{align}
\label{prop:insurance_condi}
\end{proposition}
\vskip -10mm

This result shows that the ratio between the coverage level $r$ and premium value $C_I$ depends on the average loss of performance of the system and the risk aversion of the pursuer. Under this constraint, a risk-averse buyer will have the incentive to purchase the insurance. This provides a fundamental principle for designing the insurance policy.

\subsubsection{Maximum Premium with Full Coverage}

In this section, we discuss the maximum acceptable premium the risk-averse buyer is willing to pay. According to Prop.~\ref{prop:insurance_condi}, the ratio between the coverage level and the premium $C_I/r$ is bounded by the expected and biased loss of performance of the system. The maximum premium value can be achieved when the insurer is providing full coverage as $r=1$.

\begin{proposition}\label{prop:opinsurance}
The maximum acceptable premium for the buyer is achieved under the following insurance policy:
\begin{align}
    r^*=1, \qquad C_I^* = \Phi(\hat{L}_B ).
\end{align}
\end{proposition}

Consider the PT risk aversion in (\ref{eq:prospect_theory}). Since we are considering the absolute value of losses, the utility function need to be reflected over the origin. The maximum acceptable premium can be expressed as
\begin{align}\label{eq:ci}
C_I^* =  \hat{\Phi}(\hat{L}_B) = (1-P_A^m)\cdot \lambda (\Delta U_B)^\beta.
\end{align}
        \begin{figure}[ht]
                \centering
                \includegraphics[width=0.6\textwidth]{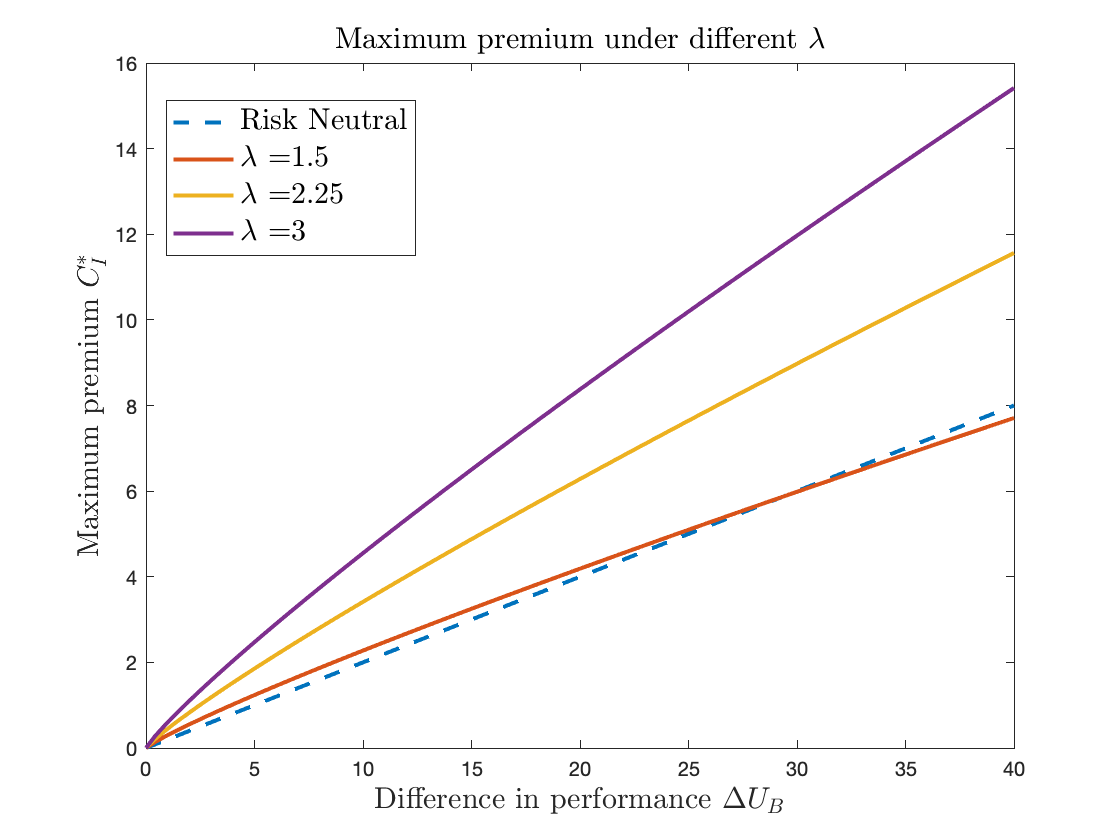}
                \caption{Maximum acceptable premium under different degrees of risk aversion.}
                \label{fig:maxci_beta}
        \end{figure}

\begin{proposition}
With full coverage $r=1$, the maximum acceptable premium is higher than the unbiased expected loss when the performance difference is relatively small, as
\begin{align*}
    C_I^*\geq \hat{L}_B \qquad \text{if} \quad 0\leq \Delta U_B \leq \lambda^{\frac{1}{1-\beta}}.
\end{align*}
\end{proposition}

We first set $P_A^m=0.8$, apply $\beta = 0.88$, $\zeta = 0.69$ in behaviour science literature and discuss the influence of loss aversion level $\lambda$ on the maximum acceptable premium $C^*_I$, which is depicted in Fig.~\ref{fig:maxci_beta}. The dotted line served as the baseline of the risk-neutral buyer, which represents the unbiased expected loss of performance. The larger value of $\lambda$ indicates that the buyer is more risk-averse against the losses. The biased loss function is concave in $\Delta U_B$ because when the $\Delta U_B$ in performance is too high, a small increase in losses has little influence on the buyer's recognition.

Risk-averse buyers are sensitive to small losses, which provides the insurer an opportunity to take advantage of the risk aversion and charge for a higher premium. From the figure, the biased expected loss is greater than the unbiased one when $\Delta U_B$ is within the tolerable range for the buyer. This range coincides with the insurance purchase constraint in Prop.~\ref{prop:insurance_condi}. If $\Delta U_B > \lambda^{\frac{1}{1-\beta}}$, we have $\Phi(\hat{L}_B)>\hat{L}_B$ and the buyer would not have the incentive to purchase cyber insurance anymore. This indicates that the insurer can increase the premium to maximum acceptable value if the buyer is going to purchase the insurance. 
        
\begin{proposition}
Cyber insurance is an incentive mechanism that encourages the buyer to have a more reliable accountability investigation.
\end{proposition}

Another key result is that cyber insurance could increase the buyer's incentive to establish a more valid accountability investigation method. As described in equation (\ref{eq:ci}), the maximum acceptable premium $C_I^*$ has a negative correlation with respect to the accountability $P_A^m$. Let $\beta = 0.88$, $\lambda = 2.25$ and $\zeta = 0.69$ as the typical values in prospect theory, the influence of accountability investigation on the maximum acceptable premium is depicted in the following figure.

        \begin{figure}[ht]
                \centering
                \includegraphics[width=0.6\textwidth]{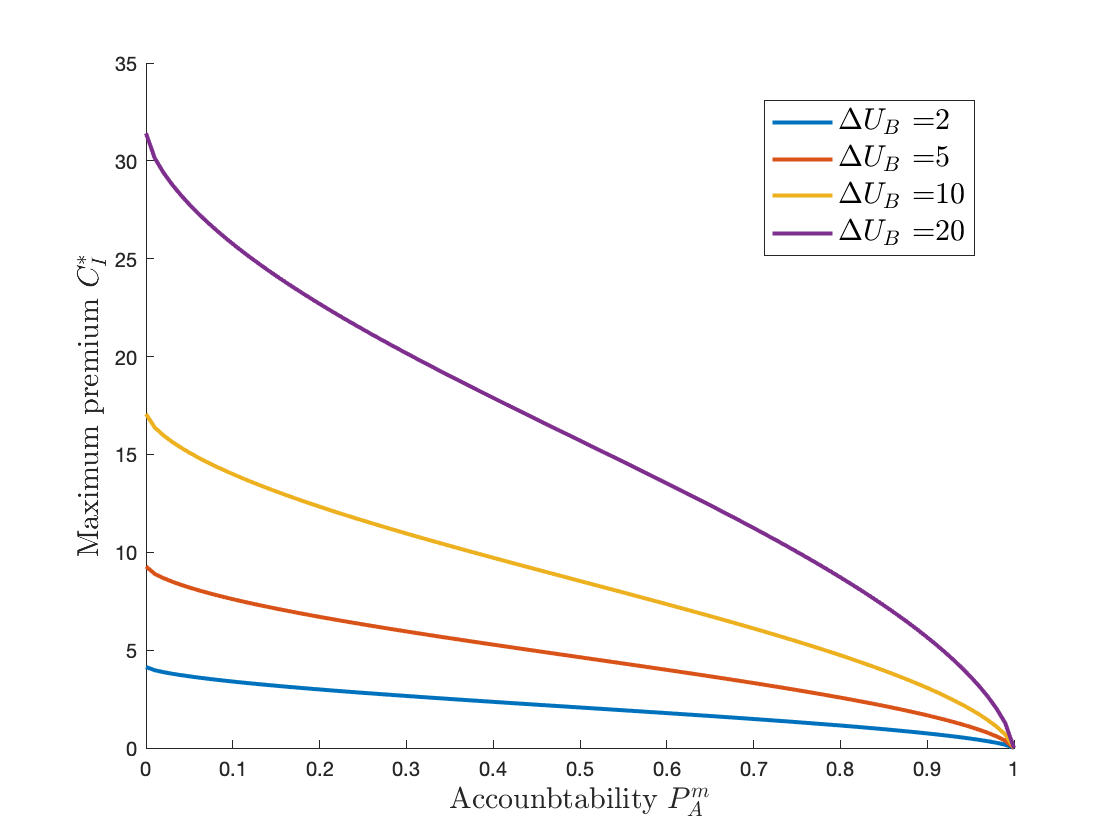}
                \caption{Relationship between accountability and maximum acceptable premium.}
                \label{fig:pa_ci}
        \end{figure}
Figure~\ref{fig:pa_ci} illustrates that a more reliable accountability investigation (larger $P_A^m$) can reduce the maximum premium of the insurance. The amount of reduction is higher if the performance differs more within two product types. If we consider the payoff function of the buyer under full insurance coverage. If the insurance company charges the maximum acceptable premium, we have
\begin{align}
    J_B(m,\delta) = C_B(m)+C^*_I.
\end{align}
The decrease in $C_I$ will reduce the total payoff $J_B$ of the buyer, resulting in a higher profit. This is the same as saying that cyber insurance provides incentives for the buyer to invest more in accountability investigation and establish a more reliable examination method to determine whether the supplier should be accountable for the incident. 

\subsubsection{Coverage Level with Given Premium}

In this section, we discuss the coverage level $r$ when the premium $C_I$ is given. As demonstrated in Prop.~\ref{prop:insurance_condi}, given a premium $C_I$, the insurance contract will be established if
\begin{align}\label{eq:r_cont}
    \frac{C_I}{\Phi(\hat{L}_B)}\leq r \leq \frac{C_I}{\hat{L}_B}.
\end{align}
This can be regarded as a constraint in the optimization problems for the buyer and the insurer. 

Given $C_I$, the buyer's problem is to find the optimal coverage level that minimizes the total payoff under insurance.
\begin{equation}
\begin{aligned}
\min_{r\in(0,1]} \quad & J_B = (1-r)\Phi(\hat{L}_B)+C_B(m)+C_I,\\
\textrm{s.t.} \quad &  \frac{C_I}{\Phi(\hat{L}_B)}\leq r \leq \frac{C_I}{\hat{L}_B}.\\
\end{aligned}
\tag{$OP_B$}
\end{equation}
Note that the buyer will make decision under biased expected loss, thus we use $\Phi(\hat{L}_B)$ in the objective function to represent her recognition. On the other hand, the insurer's problem is to find the optimal coverage level that maximizes his profit.
\begin{equation}
\begin{aligned}
\max_{r\in(0,1]} \quad & J_I = C_I - r\hat{L}_B ,\\
\textrm{s.t.} \quad &  \frac{C_I}{\Phi(\hat{L}_B)}\leq r \leq \frac{C_I}{\hat{L}_B}.\\
\end{aligned}
\tag{$OP_I$}
\end{equation}
We assume the insurer is rational and the expected loss in the objective function is unbiased.

By solving these two optimization problems ($OP_B$) and ($OP_I$), the optimal coverage levels for the buyer and the insurer are 
\begin{align*}
    r_B^* = \max\Big\{\frac{C_I}{\hat{L}_B},1\Big\}, \qquad r_I^* = \min \Big\{\frac{C_I}{\Phi(\hat{L}_B)},0\Big\}.
\end{align*}
The buyer prefers a larger coverage level at the upper bound under the constraints, while the insurer favors a lower coverage level at the lower bound. The result coincides with the fact that the insurance company and the buyer have a conflict of interest in terms of the overall payoff. However, the individual preferences of both sides need to satisfy the constraint in (\ref{eq:r_cont}) in order to establish the insurance contract in the first place.

        \begin{figure}[ht]
                \centering
                \includegraphics[width=0.6\textwidth]{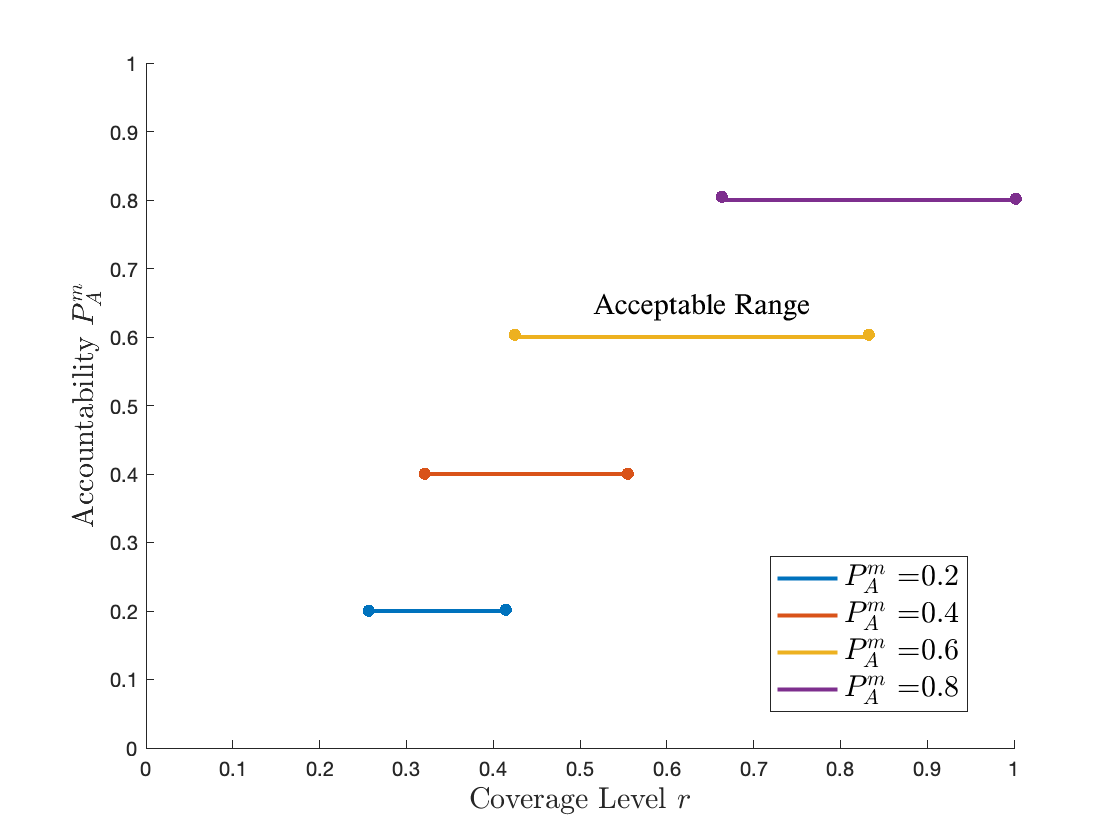}
                \caption{Coverage level under different accountability ($\Delta U_B = 6$, $C_I = 2$).}
                \label{fig:r_range}
        \end{figure}

\begin{proposition}
Given the insurance premium $C_I$, the acceptable range of coverage level $r$ will shift in the buyer's favor with larger accountability $P_A^m$.
\end{proposition}

Figure~\ref{fig:r_range} illustrates the acceptable coverage level $r$ when the performance difference $\Delta U_B = 6$ and given premium value $C_I = 2$. From the figure, both bounds of the coverage level will increase with respect to the accountability $P_A^m$. This is because both $\hat{L}_B$ and $\Phi(\hat{L}_B)$ are decreasing functions in $P_A^m$. The phenomenon shows that a more reliable accountability investigation (larger $P_A^m$) will benefit the buyer when participating in cyber insurance. Since the insurance contract will only be established under the constraint, the acceptable range of coverage level closer to $1$ will cover more portion of the losses
in the system, thus reducing the payoff that the buyer needs to pay after a system malfunction.

\subsubsection{Trade-off Between Accountability Investment and Cyber Insurance}

Lastly, we discuss the trade-off between the investment in accountability investigation and cyber insurance. From the previous discussion, a more reliable accountability investigation method (larger $P_A^m$) will reduce the maximum acceptable premium $C_I$ and increase the coverage level $r$. These will result in a more favorable insurance plan for the buyer that mitigates the losses of performance due to the supplier. However, usually, the increase in $P_A^m$ comes with a cost. This brings up the question: how much should we invest in accountability?

Suppose the cost to increase the accountability from $P_A^m$ to $P_A^{m'}$ is $C_n$. This value represents the extra funding on accountability investigation. The total payoff of the buyer before ($J_B$) and after ($J_B'$) accountability investment are
\begin{equation}
\begin{aligned}
    J_B &= (1-r)(1-P_A^m)\Delta U_B +C_B(m)+C_I\\
    J'_B &= (1-r')(1-P_A^{m'})\Delta U_B +C_B(m)+C_I'+ C_n
    \end{aligned}
\end{equation}
where $r'$ and $C'_I$ are the modified insurance plan. From previous discussion, we know that $P_A^{m'}>P_A^m$, $r'>r$ and $C_I'<C_I$. The problem is to find the optimal investment such that
\begin{align}
    J'_B - J_B\leq 0.
\end{align}
The optimal investment will depend on various factors such as the cost $C_n$, expected loss $\hat{L}_B$, the buyer's risk aversion, etc. We will illustrate the trade-off between accountability investment and cyber insurance in the following example.

\subsubsection*{Example: autonomous truck platooning}

Consider the autonomous truck platooning example in Sec.~\ref{sec:acc_test}. The accountability of the supplier takes the form
\begin{align}
            P_A^m(N)  = 1-Q\left(\frac{d}{2}+\frac{\ln(\tau)}{d}\right),
\end{align}
where $d=N^{1/2}e_d/\sigma$. Normally, the sensor difference $e_d$, supplier's reputation ratio $\tau$ and observation variance $\sigma^2$ are already given. The only variable that is completely controlled by the investigator is the number of test $N$. From the analysis in the previous section, we know that $dP_A^m/dN\geq 0$.  In order to reach a higher value of $P_A^m$, the buyer need to increase the number of tests during the investigation, which is costly in general.

Consider the insurance plan with full coverage $r=1$ and maximum premium $C_I^*$ as described in Prop.~\ref{prop:opinsurance}. We assume the buyer obeys CARA risk aversion for the expected loss. Suppose the cost to conduct one test is $c_n$. The buyer would like to find out the optimal number of tests $N$ that can minimize her payoff, which is
\begin{equation}
\begin{aligned}
\min_{N} \quad  J_B =& (1-r)\hat{L}_B+C_B(m)+C^*_I+N\cdot c_n\\
 =& C_B(m)+(1-P_A^m(N))\phi(\Delta U_B) + N\cdot c_n
\end{aligned}
\end{equation}

        \begin{figure}[ht]
                \centering
                \includegraphics[width=0.6\textwidth]{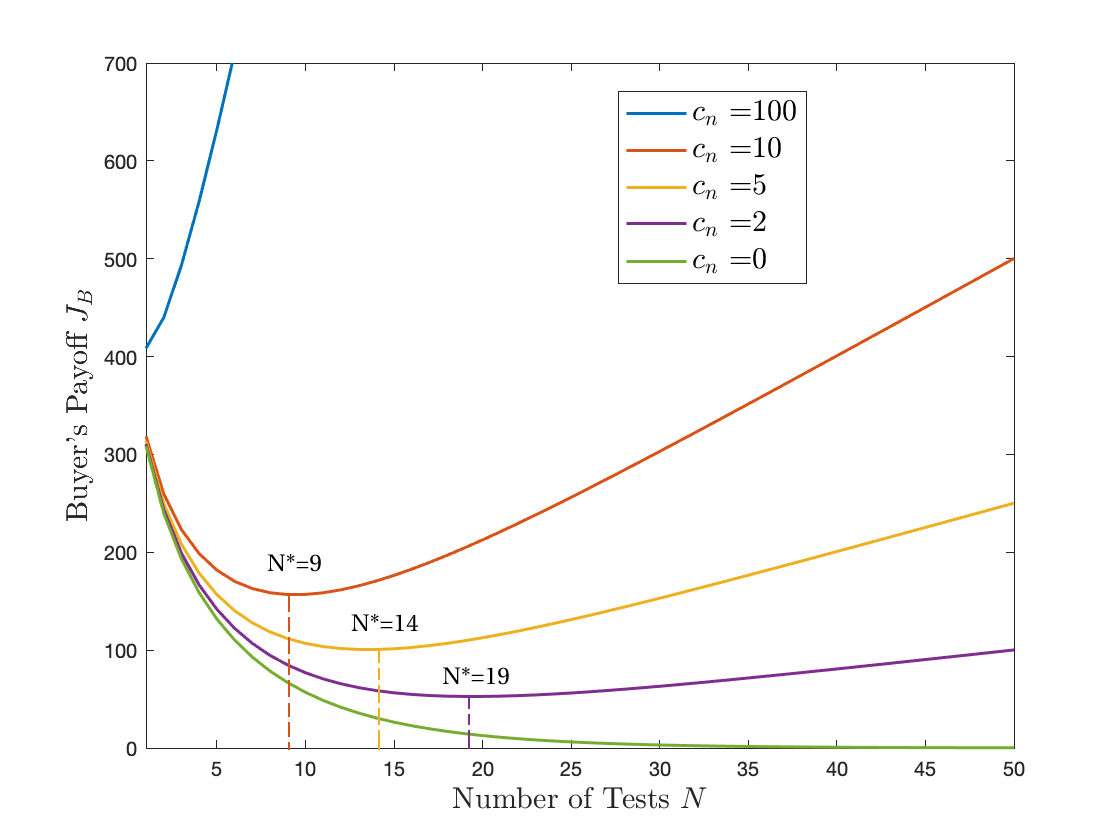}
                \caption{Optimal number of test with different test cost.}
                \label{fig:n_cost}
        \end{figure}

Figure~\ref{fig:n_cost} shows the optimal number of accountability tests with different test costs. When there is no cost to conduct one accountability test ($c_n=0$), the more test the better for the buyer. Increasing the number of tests, in general, will increase the accountability $P_A^m$. As $N\to \infty$, the accountability investigation can identify the untruthful supplier almost surely with $P_A^m\to 1$. In this case, the supplier will be penalized for the misinformation, and the payoff of the buyer will be close to zero. When the cost of each test $c_n$ increases, the optimal number of test $N^*$ will decrease. This illustrates the trade-off between accountability investigation and cyber insurance. Even though increasing the number of tests will provide a more reliable test and reduce the insurance premium, the total investment would exceed the benefit after some point, causing unnecessary payoff for the buyer. Finally, if the investigation is too costly as $c_n=100$, the buyer will never benefit from conducting an accountability investigation. It is better for the buyer to change to other comparatively low-cost investigation methods. By decreasing $c_n$, the buyer could find the optimal number of tests and achieve a lower payoff.

\section{Conclusion}

In this chapter, we have proposed a system-scientific framework to study the accountability in IoT supply chains and provided a holistic risk analysis technologically and socio-economically. We have developed stylized models and quantitative approaches to evaluate the accountability of the supplier. Two case studies have been used to demonstrate the model of accountability in the setting of autonomous truck platooning and ransomware in IoT supply chain.

We discuss the accountability investigation performance and design with a single supplier in the autonomous truck platooning case. From the parameter analysis, the reliability of the investigation can be improved with larger sensor error, more number of tests, and less observation variance. We have also showed the impact of the supplier's reputation on accountability investigation. A bad reputation will increase both accountability and wronged accountability during the investigation.

Using the smart lock case study, we have illustrated how to determine the accountability of the supplier in the IoT supply chain under a ransomware attack. A Neyman-Pearson test has been used to deal with suppliers with limited prior information. We have presented the model of the multi-stage accountability investigation with multiple suppliers in the supply chain and discussed the trade-off between detailed investigation and product replacement. 

Contract design and cyber insurance are used as economic solutions to improve the cyber resilience in IoT supply chains. By designing contracts under incentive-compatibility and individual rationality constraints, the IoT end-user can penalize the accountable supplier and reduce his incentive of providing misinformation in the first place. Cyber insurance mitigates the loss of performance by transferring the risks to a third party. We have showed that cyber insurance is an incentive-compatible mechanism that facilitate a more reliable accountability investigation from the buyer side. However, the investigator needs to balance between the accountability investment and cyber insurance to achieve a higher payoff.

\newpage
\singlespacing
\setlength\bibsep{0pt}
\bibliographystyle{abbrv}
\bibliography{acc}

\end{document}